\documentclass[manuscript]{emulateapj}

\tighten
\def\<<{{\ll}}
\def\>>{{\gg}}

\def\spose#1{\hbox to 0pt{#1\hss}}
\def\ltwig{\mathrel{\spose{\lower 3pt\hbox{$\mathchar"218$}}
     \raise 2.0pt\hbox{$\mathchar"13C$}}}
\def\gtwig{\mathrel{\spose{\lower 3pt\hbox{$\mathchar"218$}}
     \raise 2.0pt\hbox{$\mathchar"13E$}}}
\def\+/-{{\pm}}
\def\=={{\equiv}}
\def\Rstar{R_{\ast}}
\def\Mstar{M_{\ast}}
\def\Lstar{L_{\ast}}
\def\Fstar{F_{\ast}}

\def\vth{v_{th}}

\def\Mdot{\dot M}

\def\yr{{\rm yr}}

\def\solar{\odot}
\def\Msun{M_{\solar}}
\def\msbyr{\Msun/\yr}
\def\Rsun{R_{\solar}}
\def\Lsun{L_{\solar}}

\def\vinf{v_\infty}

\newcommand{\beq}{\begin{equation}}
\newcommand{\eeq}{\end{equation}}
\newcommand{\beqa}{\begin{eqnarray}}
\newcommand{\eeqa}{\end{eqnarray}}

\def\gbar{{\bar\Gamma}}

\def\sed{super-Eddington~}
\def\eff{{\mathrm{eff}}}

\def\wf{{\cal W}}

\usepackage{graphicx} 

\begin{document}

\submitted{submitted: 10 May 2004; revised 12 Jul 2004}

\title{
A Porosity-Length Formalism for 
Photon-Tiring-Limited 
\\
Mass Loss 
from Stars Above the Eddington Limit}
\author{Stanley P. Owocki}
\affil{Bartol Research Institute,
       University of Delaware,
       Newark, DE 19716}
\author{Kenneth G. Gayley}
\affil{Department of Physics and Astronomy,
       University of Iowa,
       Iowa City, IA 52242}   
\author{Nir J. Shaviv}
\affil{Racah Institute of Physics,
       Hebrew University,
        Giv'at Ram, Jerusalem 91904
	Israel}   

\begin{abstract}

We examine radiatively driven mass loss from stars near and above 
the Eddington limit.
Building upon the standard CAK theory of driving by scattering in an 
ensemble of lines with a power-law distribution of opacity, we first 
show  that the formal divergence of such line-driven mass loss as a 
star approaches the Eddington limit is actually limited by the 
``photon tiring'' associated with the work needed to lift material 
out of the star's gravitational potential.
We also examine such tiring in simple continuum-driven models in which 
a specified outward increase in opacity causes a net outward 
acceleration above the radius where the generalized Eddington 
parameter exceeds unity. When the density at this radius 
implies a mass loss too close to the tiring limit, the overall result is 
flow stagnation at a finite radius. Since escape of a net steady wind 
is precluded, such circumstances are expected to lead to extensive 
variability and spatial structure. After briefly reviewing convective and 
other instabilities that also can be expected to lead to extensive structure 
in the envelope and atmosphere of a star near or above the Eddington limit, 
we investigate how the {\it porosity} of such a structured medium can 
reduce the  effective coupling between the matter and radiation. 
Introducing a new  ``{\it porosity-length}'' formalism, we derive a simple 
scaling for the reduced effective opacity, and use this to derive an 
associated scaling for the porosity-moderated, continuum-driven mass loss 
rate from stars that formally exceed the Eddington limit. 
For a simple super-Eddington model with a single porosity length that is 
assumed to be on the order of the gravitational scale height, 
the overall mass loss is similar to that derived in previous 
porosity models,  given roughly by $\Lstar/a_{\ast} c$ 
(where $\Lstar$ is the stellar luminosity, and $c$ and $a_{\ast}$ are the 
speed of light and atmospheric sound speed). 
This is much higher than is typical of line-driven winds, 
but is still only a few percent of the tiring limit. 
To obtain still stronger mass loss that approaches observationally
inferred values near this limit,
we draw upon an analogy with the power-law distribution of line-opacity 
in the  standard CAK model of line-driven winds, and thereby introduce a 
{\it power-law-porosity} model in which the associated structure has a 
broad range of scales. 
We show that, for power indices $\alpha_{p} <1$, the mass loss rate 
can be enhanced over the single-scale model by a factor that increases 
with the Eddington parameter as $\Gamma^{-1+1/\alpha_{p}}$.
For lower $\alpha_{p}$ ($\approx 0.5-0.6$) and/or moderately large 
$\Gamma$ ($ > 3-4$), 
such models lead to mass loss rates that approach the photon tiring limit.
Together with the ability to drive quite fast outflow speeds 
(of order the surface escape speed), the derived, near-tiring-limited 
mass loss offer a potential dynamical basis to explain the observationally 
inferred  large mass loss and flow speeds of giant outbursts in 
$\eta$~Carinae and other Luminous Blue Variable stars.

\end{abstract}

\section{Introduction}

Massive, hot, luminous stars -- those of spectral types O, B, and 
WR -- continuously lose mass in strong, radiatively driven stellar winds.
For these relatively quiescent phases of mass loss (i.e., with mass loss 
rates ranging up to $10^{-5} M_{\odot}/yr$), the central 
mechanism for coupling the radiation to the outflowing gas is understood 
to be via {\it line} opacity, augmented through the systematic Doppler shift 
of the line-scattering by the velocity gradient associated with the flow 
acceleration.
The inherently nonlinear feedback between the line driving and flow 
acceleration can be solved self-consistently via a formalism first 
developed by \citet[hereafter CAK]{CAK}, with 
modern extensions providing wind solutions that are in remarkably good, 
quantitative agreement with observational inferences of key wind properties 
like the mass loss rate and wind flow speed, and with the scaling of these with 
stellar parameters like the luminosity and gravity.

However, at least some massive stars -- observationally identified as 
the so-called Luminous Blue Variables (LBV's) -- appear to undergo one or more 
phases of much stronger mass loss. Perhaps the most extreme example is 
the giant eruption of the massive LBV star $\eta$~Carinae, which is estimated to 
have cumulatively lost 2-10~$\Msun$  between 1840-1860 
\citep{DH97,Smith02}, 
representing a mass loss rate $\sim 0.1-0.5 \Msun$/yr that is about a factor 
$10,000$ times greater than can be readily explained via 
the line-driven wind formalism.
Instead, the observational association of these LBV's as being near an 
apparent upper limit in luminosity for observed stars 
\citep{HD79, HD84} 
has led to the 
general view that the strong mass loss may instead stem from the star
approaching or exceeding the so-called ``Eddington limit'', at which even the 
{\it continuum} force associated with perhaps just electron scattering 
exceeds the inward force of gravity.

But a key difficulty in understanding how the approach or breach of 
the Eddington limit might lead to stellar mass loss lies in the fact 
that both the radiative acceleration and gravity have a similar 
inverse-square scaling with radius, implying that their ratio $\Gamma$ 
(the so-called Eddington parameter) has (in simple 1-D models) a nearly 
spatially constant value throughout the star.
As such, reaching or exceeding the Eddington limit $\Gamma \ge 1$ 
would appear to leave the entire star gravitationally unbound, and so 
does not constitute an appropriate description for steady-state, 
{\it surface} wind mass loss from an otherwise stably bound central 
star.

A promising solution to this difficulty lies in relaxing the usual 
assumption of strictly one-dimensional (1-D) spherically symmetric 
stratification, and considering how the lateral structuring -- or 
``porosity''-- of a medium could lead to an effective {\em reduction} in the 
coupling between the radiation and matter \citep{Shaviv98,Shaviv00}.
In principal, this can provide a way for quasi-stationary wind outflows 
to be maintained from objects that formally exceed the Eddington limit. 
A key insight regards the fact that, in a spatially 
inhomogeneous atmosphere, the radiative transport should selectively avoid 
regions of enhanced density in favor of relatively low-density, 
porous channels between them.
This stands in contrast to the usual picture of simple 1-D, gray-atmosphere 
models, wherein the requirements of radiative equilibrium ensure that the 
radiative flux must be maintained regardless of the medium's  optical 
thickness. 
In 2-D or 3-D porous media, even a gray opacity can lead to a flux avoidance 
of the most optically thick regions, much as in frequency-dependent radiative 
transfer in a 1-D atmosphere, where the flux avoids spectral lines or 
bound-free edges that represent spectrally localized regions of non-gray 
enhancement in opacity. 
In such a case, the force opacity can be significantly smaller than the 
Rosseland mean opacity.

In a gray but porous super-Eddington medium,
the associated reduction in the coupling of the denser, more 
optically thick regions of the atmosphere can bring the 
{\it effective} Eddington parameter below unity, and thus allow for a 
stably stratified medium.
But as the density decreases outwards and the individual structures  
become more optically thin, they are again exposed to the full 
radiation. 
In a formally \sed medium, the location where the effective 
Eddington parameter exceeds unity can therefore mark the initiation of the net 
outward acceleration for a stellar wind \citep{Shaviv01MNRAS}.

In principal, the full solution for a wind outflow in this case 
requires self-consistent dynamical solution of both the formation of the 
lateral structure, as well as multi-dimensional transport of the 
radiation to determine the effective radiative acceleration throughout 
the thick to thin regions of the medium. 
As this represents a quite challenging and computationally intensive 
effort, it seems appropriate first to seek the development of simpler, perhaps 
phenomenological approaches that can provide some basic insights into 
the key properties of wind structure and the likely scalings of the 
resultant mass loss rates and flow speeds with relevant parameters.
Moreover, phenomenological approaches have the advantage that their results 
can more easily be implemented in more complex systems. 

The aim of this paper is to offer such a simplified, heuristic 
approach, one that takes particular advantage of insights and analogies 
from the much more well-established theory for line-driven winds.
In the following we thus begin (\S 2) with a general overview of the basic 
properties and scalings of the standard CAK wind formalism for driving 
by a power-law ensemble of spectral lines.
We next (\S 3) discuss how the loss of radiative energy, or ``photon 
tiring'', places a fundamental limit on both line- and continuum-driven 
mass loss, and then (\S 4) review how a super-Eddington condition in 
the deeper layers of star can instead lead to convection, pressure 
inversion, and generally a highly structured medium.
To account for the associated porosity reduction in the radiative 
driving, we next  (\S 5) introduce a simple clump-absorption 
formalism that, in conjunction with a new ``porosity length'' {\it 
ansatz} (somewhat analogous to the mixing length formalism for 
convective energy transport) allows the 
estimation of the porosity-moderated mass loss rate.
Drawing on analogies with the CAK model for line-driving, we then 
(\S 6) develop a {\em power-law} porosity formalism and apply this to derive 
scalings for the mass loss rate and solutions for the wind velocity law.
A discussion section (\S 7) then compares these scalings with previous 
porosity models, and with inferred observational properties of the 
giant eruption in $\eta$~Carinae,
including its bipolar form.
We conclude (\S 8) with a general summary and outlook for future work.

\section{Radiatively Driven Mass-Loss by Line and/or Continuum Opacity} 

\subsection{General Equation of Motion}

Consider a steady-state stellar wind outflow in which the net acceleration
$v (dv/dr)$ in the radial flow speed $v(r)$ at radius $r$ results from 
a radiative acceleration $g_{rad}$ that overcomes the inward gravitational 
acceleration, $G\Mstar/r^{2}$,
\beq
v {dv \over dr} 
=- \frac {G\Mstar} {r^2} \; + \; g_{rad} \, - {1 \over \rho} {dP \over 
dr} \, ,
\label {dimeom}
\eeq
where the mass density $\rho$ and the gas pressure $P = \rho a^{2}$ 
are related (by the perfect gas law) through the isothermal sound 
speed $a$.
Using the steady-state equation of mass continuity ($\rho v r^{2} =$ 
constant) to eliminate the density $\rho$, 
eqn.~(\ref{dimeom}) takes the form
\beq
\left [ 1 - {a^{2} \over v^{2} } \right ] \, v {dv \over dr} 
=- \frac {G\Mstar}{r^2} \; + \; g_{rad} \, + {2 a^{2} \over r } 
- {d a^{2} \over dr } \, ,
\label {eom}
\eeq
where the terms containing the isothermal sound speed $a$ arise from the gas 
pressure gradient.
The square-bracket factor on the left-hand-side (LHS) allows for a smooth mapping 
of the wind base onto a hydrostatic atmosphere below the sonic point, where $v<a$.
But in radiatively driven winds the pressure terms on the right-hand-side (RHS) 
are  generally negligible since, compared to the gravitational acceleration term
that must be overcome to drive a wind, these are of order 
$w_{s} \equiv (a/v_{esc})^{2} \approx 0.001$, where 
$v_{esc} \equiv \sqrt{2G\Mstar/\Rstar}$ is the escape 
speed from the stellar surface radius $\Rstar$. 
In the development here, we consequently drop these RHS gas 
pressure  terms, but for now  still retain the LHS factor to allow the option 
of deriving wind solutions that map smoothly onto a hydrostatic atmosphere.

Since the key to a stellar wind is to overcome gravity, it is 
convenient to define a dimensionless equation of motion that scales all 
the accelerations by gravity,
\beq
(1-w_{s}/w) \, w' = -1 + \Gamma_{rad} 
\, ,
\label{dimlesseom}
\eeq
where $\Gamma_{rad} \equiv g_{rad} \, r^{2}/G\Mstar$, and
the gravitationally scaled inertial acceleration is
\beq
w' \equiv \frac
{r^2 v dv/dr}{G\Mstar }\, .
\label{wp-def}
\eeq
In terms of an inverse radius coordinate $x \equiv 1-\Rstar/r$,
note that $w' = dw/dx$, where $w \equiv v^{2}/v_{esc}^{2}$ represents the 
ratio of wind kinetic energy to the gravitational binding 
$v_{esc}^{2}/2 \equiv G\Mstar/\Rstar$ from the stellar surface radius $\Rstar$.

For isotropic opacity the dimensional radiative acceleration is 
set by integration over frequency $\nu$ of the star's radiative 
flux $F_{\nu}$ weighted by the opacity $\kappa_{\nu}$
\beq
g_{rad} = \int_{0}^{\infty} \, d\nu \, \kappa_{\nu} F_{\nu}/c \, .
\label {graddef}
\eeq
In the common, simple case of a continuum dominated by free electron 
scattering, the opacity  $\kappa_{e}$ ($= 0.2 (1+X) \, cm^{2}/g$ for fully 
ionized plasmas with hydrogen mass fraction $X$, which we take here 
to have the standard cosmic value $X \approx 0.7$.) is strictly gray 
(frequency independent), allowing it to be pulled out of the 
frequency integration. 
This yields an associated acceleration
\beq
g_{e} = \kappa_{e} F/c = { \kappa_{e} \Lstar \over 4 \pi r^{2} c } \, ,
\label {gedef}
\eeq
where $\Lstar$ is the star's bolometric luminosity.
In the associated gravitationally scaled acceleration, the 
inverse-radius-squared dependence of the flux cancels with that from gravity, 
yielding the so-called Eddington parameter
\beq
\Gamma_{e} = { \kappa_{e} \Lstar \over 4 \pi G \Mstar c } 
\, ,
\label {gamedd}
\eeq
which for 1-D radiative transport in the outer envelope and surface 
layers is a  {\it spatial constant}, set by the ratio $\Lstar/\Mstar$ of 
stellar luminosity to mass.

This lack of spatial modulation presents a key difficulty for driving 
a steady wind mass loss by electron scattering,
since a wind requires an outward acceleration that transitions from 
less than gravity in the interior to above gravity in the outflowing 
surface layers.
As we discuss below (\S 5 \& 6), development of lateral structure can lead 
to a porosity moderation of continuum driving that allows such a
quasi-steady mass loss.

\subsection{CAK Formalism and Scalings for Line-Driven Stellar Winds}

The most well-established formalism for radiative driving in massive stars 
invokes the operation of {\it line} opacity from bound-bound transitions, 
as first worked out in detail by CAK.
As resonant processes, bound transitions have inherently large cross-sections, 
but only for radiation within a narrow frequency band of width 
$\Delta \nu \ll \nu_{l}$ centered on some line-center resonance 
frequencies $\nu_{l}$.
In the dense layers of a star, the diffusion of radiative energy means 
that the flux is reduced in proportion to the enhanced opacity, 
keeping the effective line force small (see eqn.~\ref{graddef}).
But within the outflowing wind, the Doppler shift of the 
line-resonance out of the absorption shadow of underlying material 
exposes the line opacity to a less attenuated flux, and so can lead 
to a strong radiative force. 

Consider for example the case of a single isolated line of resonance frequency 
$\nu_{l}$ within the stellar flux spectrum $F_{\nu}$, with 
integrated strength $\kappa_{l} \Delta \nu/\nu_{l} \equiv q \kappa_{e}$
over frequency width $\Delta \nu$.
For the simple case of radially streaming radiation in a supersonic 
flow, the use of approximations introduced by Sobolev (1960) gives the ratio 
of the line acceleration to gravity the scaling
\beq
\Gamma_{q} =  
W_{\nu}
\Gamma_{e} \, { 1 - e^{-q t_{e} } 
\over t_{e} } , 
\label {gsingline}
\eeq
where 
$W_{\nu} = \nu_{l} F_{\nu}/F$ is a line flux weight, 
$q$ is the dimensionless measure of line opacity, and
$t_{e} \equiv \rho \kappa_{e} c / (dv/dr)$ is the radial ``Sobolev optical 
depth'' of a line with $q=1$.
An important aspect of this form is that it allows evaluation of the 
radiative acceleration in terms of strictly {\it local} quantities, 
namely the density $\rho$, velocity gradient $dv/dr$, and line 
strength $q$.

A key advance of the CAK analysis was to introduce a formalism for 
including the cumulative effect of a large number of lines of varying 
strengths $q$, under the simplifying assumptions that these are independent.
The central ansatz is that the  flux-weighted number distribution 
of lines is a power law in line-strength
\beq
q { dN \over dq } = {1 \over \Gamma (\alpha) } \, 
\left [ { q \over {\bar Q} } \right ]^{\alpha - 1} \, ,
\label {dndqcak}
\eeq
where $\Gamma (\alpha)$ is the complete Gamma function, and  $\alpha$ is the 
CAK exponent.
Here the line normalization ${\bar Q}$ is related to the usual CAK $k$ parameter 
through $k= {\bar Q}^{1-\alpha} \, \left ( \vth/c \right)^{\alpha}/(1-\alpha)$;
it offers the advantages of being a dimensionless
measure of line-opacity that is independent of the assumed ion thermal
speed $\vth$,  with a nearly constant characteristic value of order
${\bar Q} \sim 10^3$ for a wide range of ionization conditions 
\citep{Gayley95}.

Integrating the single-line force (\ref{gsingline}) over the 
distribution (\ref{dndqcak}), we obtain for the ratio of the CAK line force 
to gravity
\beq
\Gamma_{CAK}
= { {\bar Q} \Gamma_{e} \over (1-\alpha) ({\bar Q} t_{e} )^{\alpha} } \, 
\equiv C (w')^{\alpha}.
\label {gamcak}
\eeq
In the latter definition, we have eliminated the density $\rho$ 
(within $t_{e}$) in favor of the mass loss rate $\Mdot=4 \pi r^2 \rho v$ 
for the assumed steady, spherical expansion, with the line-force constant 
defined by
\beq
C \equiv
\frac {1}{1-\alpha} \, \left [ \frac {\Lstar} {\Mdot c^2} \right ]^\alpha \,
\left [ {\bar Q} \Gamma_{e} \right ]^{1-\alpha} \, .
\label {cdef}
\eeq
Note that, for fixed sets of parameters for the star ($\Lstar$, $\Mstar$, 
$\Gamma_{e}$) and
line-opacity ($\alpha$, $\bar Q$), this constant scales with the mass loss rate 
as $C \propto 1/\Mdot^{\alpha}$.

As already noted, the smallness of the dimensionless sound-speed 
parameter $w_{s} \approx 0.001 \ll 1 $ implies that gas pressure plays 
little role in the dynamics of any radiatively driven stellar wind. 
Hence to a good approximation, we can obtain accurate solutions by 
analyzing the much simpler limit of vanishing sound speed 
$a \propto \sqrt{w_{s}} \rightarrow 0$, for which the line-driven-wind
equation of motion reduces to
\beq
 w' + 1 - \Gamma_{e} = C (w')^\alpha .
\label{cakeom}
\eeq
Since the parameters $\Gamma_{e}$ and $C$ are spatially constant, the 
solution is independent of radius.
For high $\Mdot$ and so small $C$ there are no solutions,
while for small $\Mdot$ and high $C$ there are two solutions.
The CAK critical solution corresponds to a {\it maximal} mass loss rate, and
requires a {\it tangential} intersection between line-force and
combined inertia plus gravity, for which
\beq
\alpha \, C_c  {w'_c}^{\alpha-1}=1 \, ,
\label {wder}
\eeq
and thus
\beq
w'_c= (1- \Gamma_{e}) \frac {\alpha}  {1-\alpha} \, ,
\label {weq}
\eeq
with
\beq
C_c= \frac {1} {\alpha^\alpha} \, 
\left [ { 1-\Gamma_{e}  \over 1-\alpha } \right ]^{1 - \alpha} \, .
\label{c-def}
\eeq
Using eqn.~(\ref{cdef}), 
this then yields the standard CAK scaling for the mass loss rate
\beq
\Mdot_{CAK}=\frac {\Lstar}{c^2} \; \frac {\alpha}{1-\alpha} \;
{\left[ \frac {\bar {Q} \Gamma_{e}}{1- \Gamma_{e}} \right]}^{{(1-\alpha)}/{\alpha}} \, .
\label{mdcak}
\eeq
Moreover, since the scaled equation of motion (\ref{cakeom}) has no explicit 
spatial dependence, the scaled critical acceleration $w'_c$ applies throughout 
the wind.
This can therefore be trivially integrated to yield 
\beq
w(x) = w(1) \, x
\label {wcakvlaw}
\eeq
where $w(1) = (1-\Gamma_{e}) \alpha/(1-\alpha)$ is the terminal value of 
the scaled flow energy.
In terms of dimensional quantities, this represents
a specific case of the general ``beta''-velocity-law,
\beq
v(r)=
v_\infty
\left ( 1- \frac {\Rstar}{r} \right )^{\beta} \, ,
\label{CAK-vlaw}
\eeq
where here $\beta=1/2$, and
the wind terminal speed $v_\infty = v_{esc} 
\sqrt{\alpha(1-\Gamma_{e})/(1-\alpha)}$ scales with
$v_{esc} \equiv \sqrt{2G\Mstar/\Rstar}$, the escape speed from the
stellar surface radius $\Rstar$.

\subsection{Extensions of the Basic CAK Theory}

Modern implementations of CAK theory have included refinements to 
account for proper integration of the finite-angle stellar disk 
\citep{FA86,PPK86}, and ionization state 
variations with radius \citep{Abbott80,Pauldrach87,Kudetal89}.
Together with corrections for a non-zero gas pressure  
\citep[see appendix of][]{OU04}, 
these typically only introduce order-unity corrections to the above theory, 
and so the scalings derived here still form a good general base to 
compare with continuum-driven formalism we develop below.

Another important extension regards the role of multi-line scattering, 
which occurs whenever the velocity separation between optically thick 
lines obeys $\Delta v < \vinf $. 
Analyses by \citet{FC82} and \citet{GOC95}
show, however, that the same CAK scalings still roughly apply as 
long as the spectral distribution of lines is uniform (i.e., follows 
Poisson statistics).
The total wind momentum then scales as $\Mdot \vinf \approx 
(\vinf/\Delta v) \Lstar/c$, thus providing a potential line-driven model 
for the dense winds of WR stars, which are generally inferred to exceed 
the single scattering limit $\Mdot \vinf = \Lstar/c$, sometimes by a factor of 
10 or more.

Finally, because of its relevance to our development below of a 
continuum-driving model, we note in passing another 
modest conceptual modification to classic CAK theory, namely to account for 
the fact that, in any discrete distribution of lines, the power-law number 
approximation should be truncated at a line-strength corresponding the 
strongest, discrete line.
Using an exponential truncation at a line strength $q_{max} = 
{\bar Q}$ (at which $q (dN/dq) \approx 1$), 
the modified, gravity-scaled CAK line force becomes
\beq
\Gamma_{lines}
= 
\Gamma_{CAK}
\left [ \left ( 1+1/{\bar Q} t_{e} \right )^{1-\alpha} - 
\left ( 1/{\bar Q} t_{e} \right )^{1-\alpha} \right ] \, .
\label {gamqmax}
\eeq
In luminous stars with $\Gamma_{e}$ within a factor few of unity,
the scaled CAK line force is of order 
$\Gamma_{CAK} \approx 1-\Gamma_{e}$, implying that 
${\bar Q} t_{e} \approx  ({\bar Q} \Gamma_{e}/(1-\Gamma_{e}))^{1/\alpha} \gg 1$, 
since typically ${\bar Q} \approx 10^{3}$. 
In such cases, the square bracket correction in eqn.~(\ref{gamqmax}) is 
therefore very nearly unity.
For lower luminosity stars, however, this square-bracket corrects the 
CAK line-force scaling to impose an overall upper limit 
$\Gamma_{lines} \le {\bar Q} \Gamma_{e}$.
Since a wind requires the driving force to exceed gravity, 
$\Gamma_{lines} > 1$, this shows that normal line-driven mass loss is only 
possible for stars with $\Gamma_{e} > 1/{\bar Q}$.

\section{``Photon Tiring'' as a Fundamental Limit to Mass Loss}

\subsection{Line-Driven Winds near the Eddington Limit}

As a star approaches the classical Eddington limit $\Gamma_{e} 
\rightarrow 1 $, the standard CAK scalings predict the mass loss rate 
to diverge as $\Mdot \propto 1/(1-\Gamma_{e})^{(1-\alpha)/\alpha}$, but 
with a vanishing terminal flow speed $\vinf \propto \sqrt{1-\Gamma_{e}}$.
The former might appear to provide an explanation for the large mass 
losses inferred in LBV's, but the latter fails to explain the 
moderately high inferred ejection speeds, e.g. the  500-800 km/s 
kinematic expansion inferred for the Homunculus nebula of $\eta$~Carinae
\citep{Smith02}.

Moreover, of course, such a divergence of the mass loss rate is 
precluded by the finite energy available in the stellar luminosity $\Lstar$, 
which sets a so-called  ``photon-tiring'' limit for lifting mass out of 
the gravitational potential from the stellar surface
\citep{OG97},
\beq
\Mdot_{tir} = { \Lstar \over v_{esc}^{2}/2 } 
= { \Lstar \over G\Mstar/\Rstar } 
= 0.032 \, {\Msun \over {\rm yr}} ~ L_{6} ~ {\Rstar/\Mstar \over \Rsun/\Msun} 
\, ,
\label{mdtir}
\eeq
where $L_{6} \equiv \Lstar/10^{6} \Lsun$.
Comparison with eqn.~(\ref{mdcak}) shows that photon tiring would
limit CAK winds whenever
\beq
1-\Gamma_{e} < {\bar Q} \Gamma_{e} 
\left [ {\alpha \over 1-\alpha} \, { v_{esc}^2 \over 2 c^{2} } 
\right ]^{\alpha/(1-\alpha)} \, .
\label {caktirlim}
\eeq
For typical parameters ${\bar Q} \approx 2000$ and 
$v_{esc}^{2}/2c^{2} \approx 10^{-5}$, we find that for $\alpha=2/3$ 
photon tiring does not become important until the star is {\it very} 
close to the Eddington limit,
\beq
1-\Gamma_{e} < 2000
\left [2  \times 10^{-5} \right ]^2  = 8 \times 10^{-7} ~~ ; ~~ \alpha=2/3 \, .
\label {a2b3tir}
\eeq
However, for just a somewhat smaller CAK power index $\alpha=1/2$, the 
condition is {\it much} less stringent,
\beq
1-\Gamma_{e} < 2000
\left [ 10^{-5} \right ]  = 2 \times 10^{-2} ~~ ; ~~ \alpha=1/2 \, .
\label {a1b2tir}
\eeq
This emphasizes that the CAK mass loss rate is extremely sensitive to 
the power index $\alpha$, 
particularly near the Eddington limit, where the term within the 
square bracket in eqn.~(\ref{mdcak}) has a large numerical value, 
which is then raised to a power that depends on $\alpha$.

\subsection{Photon Tiring in Continuum-Driven Mass Loss}

Let us now examine photon tiring for continuum-driven flows, ignoring 
the effect of line-opacity, and simply considering that the continuum opacity
now has a known radial dependence $\kappa_{c} (r) $.
Specifically, we assume this opacity
{\it increases outward}
until some layer  becomes super-Eddington, i.e., 
$\Gamma_{c} > 1$, where now 
$\Gamma_{c} (r) \equiv \kappa_{c} (r) \Lstar/4\pi G \Mstar c$
is the generalized continuum Eddington parameter.
Defining for convenience the radius $r=\Rstar$ to be the radius at 
which $\Gamma_{c}(\Rstar)=1$, let us assume moreover that this 
represents the sonic point of an initiated steady-wind outflow.
The density $\rho_{\ast}$ and sound speed $a_{\ast}$ at this point set the
mass loss rate $\Mdot = 4 \pi \Rstar^2 \rho_{\ast} a_{\ast} $,
but otherwise gas pressure terms have negligible effect 
in the further supersonic acceleration of the outflow.

In terms of the scaled wind energy $w$ and scaled inverse radius 
$x$ defined in \S 2.1, eqn.~(\ref{dimlesseom}) again represents the scaled 
equation of motion, with $\Gamma_{rad} = \Gamma_{c} (x)$.
Note, however, that in this form in which $\Gamma_{c}(x)$ is presumed to be
a known spatial function, the mass loss rate itself has scaled out,
so that the resulting velocity law would be entirely independent
of the amount of mass accelerated. 

More realistically, as noted above,
a given radiative luminosity can only accelerate a limited mass loss rate
before the energy expended in accelerating the outflow against gravity would
necessarily come at the expense of a notable reduction in the radiative
energy flux itself.  
To take account of this photon tiring, we must
reduce the  radiative luminosity according to the gained kinetic and potential 
energy of the flow,
\beq
L(r) =  \Lstar  - 
\Mdot  
\left [ { v^2 \over 2 } + {G\Mstar \over \Rstar } - {G\Mstar \over r } \right ] 
\, ,
\eeq
which implies for the tiring-corrected  Eddington parameter that 
\beq
    \Gamma_{rad} (x) = \Gamma_{c} (x) \left [ 1 - m (w+x) \right] \, ,
\label{gradtir}
\eeq
where the gravitational ``tiring number'',
\beq
m \equiv {\Mdot \over \Mdot_{tir}}
={ \Mdot G\Mstar \over \Lstar \Rstar } 
  \approx 0.012 \, { \Mdot_{-4} V_{1000}^2 \over L_6 } 
\, ,
\label{mtirdef}
\eeq
characterizes the fraction of radiative energy lost in lifting
the wind out of the stellar gravitational potential from $\Rstar$.
The last expression allows easy evaluation of the likely importance of 
photon tiring for characteristic scalings, 
where $\Mdot_{-4} \equiv \Mdot/10^{-4}\, \Msun/yr$, 
$L_6 \equiv \Lstar/10^6 \,\Lsun$, 
and $V_{1000} \equiv v_{esc}/1000\, km/s $
$\approx 0.62 \,(\Mstar/\Rstar)/(\Msun/\Rsun)$.

Applying the tiring-corrected Eddington parameter eqn.~(\ref{gradtir}) into 
the dimensionless equation of motion (\ref{dimlesseom}), we find
\beq
\left ( 1- { w_{s} \over w} \right) \, 
{dw \over dx} =
-1 + \Gamma_{c} (x) [1 - m (w+x)] ,
\label{eomtir}
\eeq
where for typical hot-star atmospheres the sonic-point boundary value 
is very small, $w(0) = w_{s} = w_{\ast} \equiv a_{\ast}^2 \Rstar/2G\Mstar < 10^{-3}$.
Through most of the wind acceleration, i.e., except near the sonic point itself, 
the sound-speed term with $w_{\ast}$ can thus be neglected.
In the idealized limit $w_{\ast} \rightarrow 0$, we can use
integrating factors to obtain an explicit solution
to $w(x)$ in terms of the integral quantity 
${\bar \Gamma_{c}} (x) \equiv \int_0^x dx' \Gamma_{c}(x')$,
\beq
w(x) = -x + {1 \over m } \, \left [ 1 - e^{-m {\bar \Gamma_{c}} (x) } \right ] 
\, .
\label{wtirsoln}
\eeq

\begin{figure}
\plotone{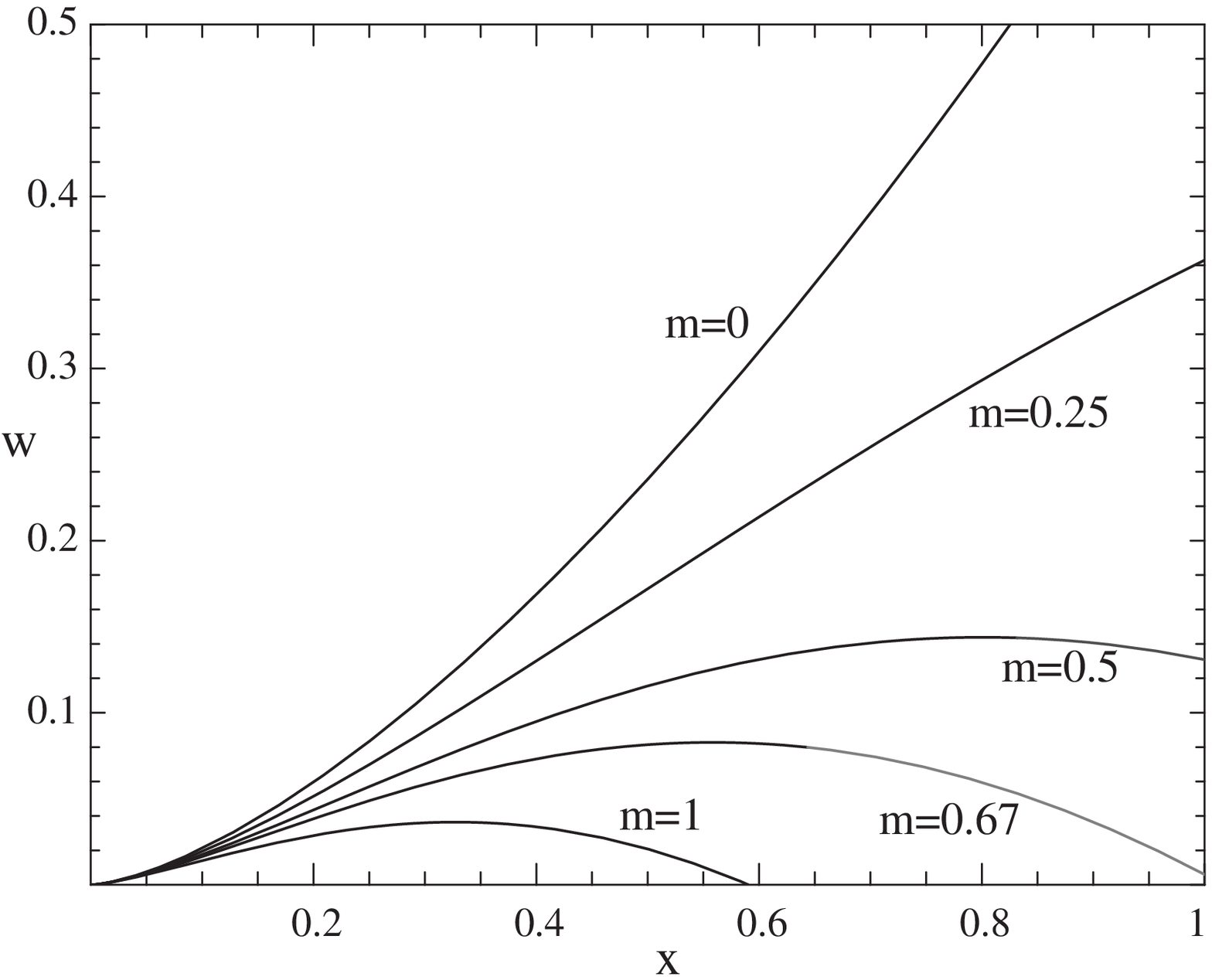}
\plotone{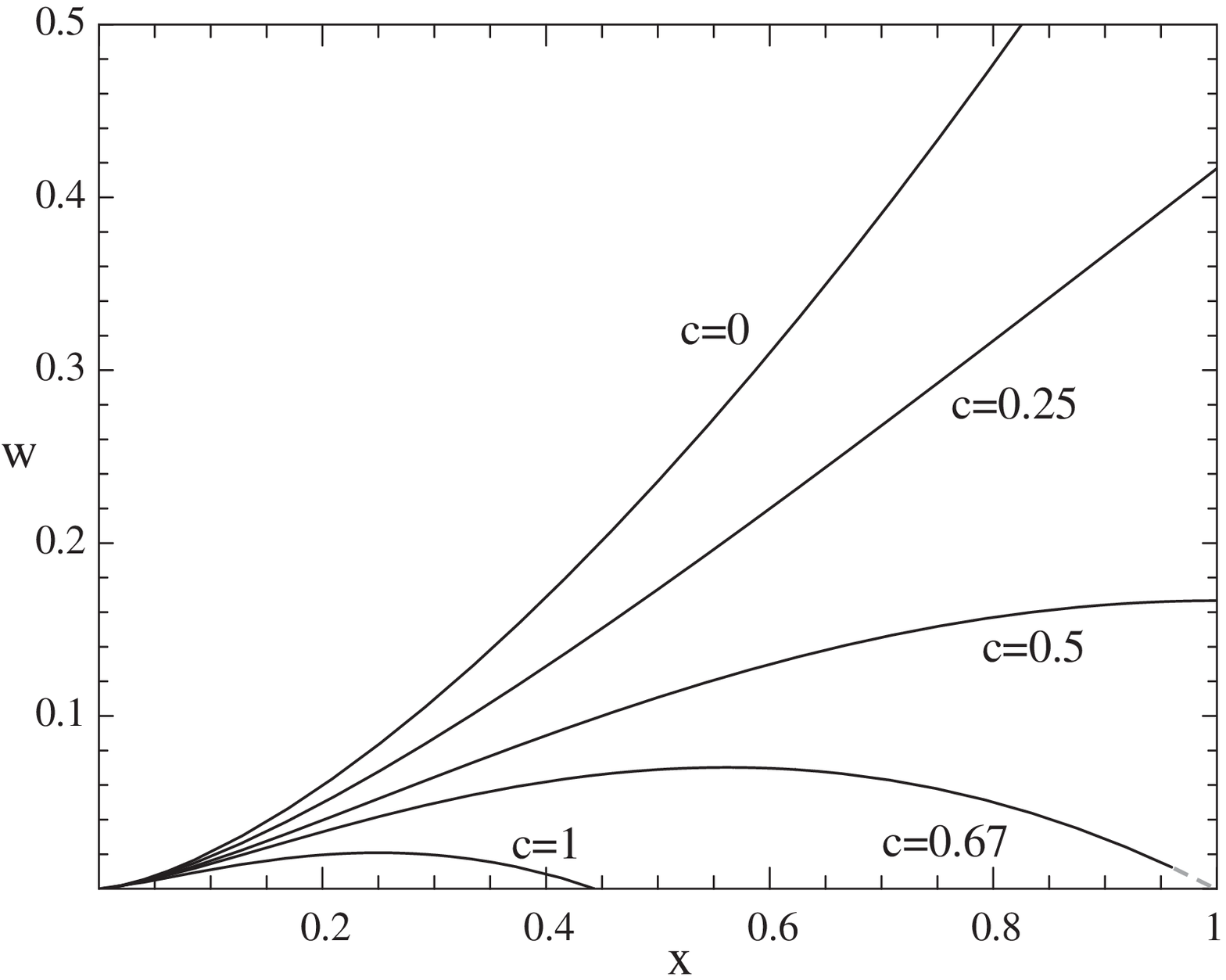}
\caption{
a. Wind energy $w$ vs.~scaled inverse radius $x$($\equiv 1-\Rstar/r$),
plotted for Eddington parameter $\Gamma_{c} (x) = 1 + \sqrt{x}$
with various photon tiring numbers $m$.
b. Same as (a), except for weak tiring limit $m \ll 1$, and for 
various constants $c$ in the Eddington parameter scaling
$\Gamma_{c}(x)=1+\sqrt{x}-2c x$.
}
\end{figure}

As a simple example, consider the case\footnote{The choice of these 
functions is arbitrary, to illustrate the photon-tiring effect within 
a simple model. More physically motivated models based on a medium's  
porosity are presented in \S~6.4.} with
$\Gamma_{c} (x) = 1 + \sqrt{x}$, for which ${\bar \Gamma_{c}} = x + 2x^{3/2}/3$.
Fig.~1a plots solutions $w(x)$ vs.~$x$ from eqn.~(\ref{wtirsoln})
with various $m$.
For low $m$, the flow reaches a finite speed at large radii ($x=1$), 
but for high $m$, it curves back, stopping at some finite {\it stagnation}
point $x_s$, where $w(x_s) \equiv 0$.  
The latter solutions represent flows for which the mass
loss rate is too high for the given stellar luminosity to be able 
to lift the material to full escape at large radii.
By considering the critical case $w(x=1)=0$,
we can define a maximum mass loss rate from $m_{max}$,
given from eqn.~(9) by the transcendental relation,
\beq
m_{max} = 1 - e^{-m_{max} {\bar \Gamma_{c}} (1)}
\approx 1-e^{2-2 {\bar \Gamma_{c}} (1)} 
,
\eeq
where the last expression provides a simple explicit 
approximation\footnote{This has a maximum error of ca. 8\%. 
A somewhat more complex, explicit approximation is 
$m_{max} = 1-\exp[5(1-{\bar \Gamma_{c}}(1)^{2/5})]$, 
which is accurate to within a half percent. 
}
for any realistic ${\bar \Gamma_{c}} (1) >1$.
Note that regardless of how large ${\bar \Gamma_{c}} (1) $ becomes, 
it is always true that $m_{max}<1$, simply reflecting the fact that 
the mass loss is always limited by the rate at which the radiative luminosity 
can lift material out of the gravitational potential from $\Rstar$. 
For a given photon tiring number $m<1$, the minimum integrated Eddington 
parameter required to ensure full escape from the potential is
\beq
{\bar \Gamma_{c} } (1) = { - \ln (1-m) \over m}  
\, .
\eeq

Even without photon tiring, a similar stagnation can occur from an 
outward reduction in the radiative driving.
In the limit of negligible tiring $m \ll 1$, the flow solution 
(\ref{wtirsoln}) 
simplifies to
\beq
w(x) \approx {\bar \Gamma_{c}} (x) - x .
\eeq
For a limited super-Eddington domain, 
the critical case of marginal escape with zero terminal velocity,
$w(1) = 0$, is now set in general by the condition ${\bar \Gamma_{c}} (1) = 1$.
For example, consider the specific case of a nonmonotonic 
$\Gamma_{c} (x) = 1 + \sqrt{x} - 2c\, x$,
for which then ${\bar \Gamma_{c}} (x) = x + 2x^{3/2}/3 - c x^{2}$.
Fig.~1b plots results for various $c$.
For all ${\bar \Gamma_{c}} (1) < 1$ (i.e., $c>2/3$), the material stagnates 
at the radius where ${\bar \Gamma_{c}} (x_s) = x_s = 4/9c^{2}$, 
and so cannot escape the system in a steady-state flow.
In a time-dependent model, such material can be expected to accumulate
at this stagnation radius, and possibly eventually fall back to the star.
This represents another way in which, instead of a steady outflow,
a limited super-Eddington region could give rise to an extended envelope with
either a mass circulation or a density inversion.

\section{Stellar Envelope Consequences of Breaching the Eddington Limit}

\subsection{Convective Instability of Deep Interior}

It should be emphasized that locally exceeding the Eddington limit
need {\it not} necessarily lead to initiation of a mass outflow.
As first shown by \citet{JSO73}, in the stellar envelope 
allowing the Eddington parameter $\Gamma \rightarrow 1$ 
generally implies through the Schwarzschild criterion that material 
becomes {\it convectively unstable}.
Since convection in such deep layers is highly efficient, the radiative
luminosity is reduced, thereby lowering the associated radiative Eddington
factor away from unity.
%

This suggests that a radiatively driven outflow should only be initiated
{\it outside} the region where convection is efficient.
An upper bound to the convective energy flux is set by
\beq
F_{conv} \approx v_{conv} \, l \, dU/dr \ltwig a \, H \, dP/dr \approx a^3 \rho , 
\eeq 
where $v_{conv}$, $l$, and $U$ are the convective velocity, mixing length,
and internal energy density, and $a$, $H$, $P$, and $\rho$ are the sound speed,
pressure scale height, pressure, and mass density.
Setting this maximum convective flux equal to the total stellar energy 
flux $\Lstar/4 \pi r^2$ yields an estimate for the maximum mass loss rate 
that can be initiated by radiative driving,
\beq
\Mdot \le {\Lstar \over a^2 } \equiv \Mdot_{max,conv} = 
{ v_{esc}^{2} \over 2 a^{2} } \Mdot_{tir} \, ,
\eeq
where the last equality emphasizes that, for the usual case of a sound
speed much smaller than the local escape speed, $a \ll v_{esc}$, such
a mass loss would generally be well in excess of the photon-tiring
limit set by the energy available to lift the material out of the
star's gravitational potential (see eqn.~\ref{mdtir}).  
In other words, if a wind were to originate from where convection becomes
inefficient, the mass loss would be so large that it would  use all the 
available luminosity to accelerate out of the gravitational potential.

A central conclusion here is accordingly that, while convective transport 
provides an alternative to a super-Eddington condition in the deep interior, 
it cannot be the regulation mechanism that would allow for a 
smooth transition to a steady wind mass loss of the near-surface layers.

\subsection{Hydrostatic Pressure Inversion in a Super-Eddington Layer}

Even above the inefficient-convection radius, a {\it limited} 
super-Eddington domain could, instead of an outflow, 
merely induce a pressure inversion layer \citep{Maeder89},
set by integrating the equation of hydrostatic equililbrium 
(cf.\  eqn.~\ref{dimeom})
\beq
{d \ln P \over dr} =  { \Gamma -1  \over H} \,
\label{pinv}
\eeq
where $H \equiv a^2 r^2/G\Mstar$ is the usual gravitational scale height.
As an example, for a narrow ($\Delta r \ll r$), isothermal, 
super-Eddington layer,
the pressure would increase by a factor exp$[(\Delta r/H)(\gbar - 1)]$,
where $\gbar$ is the average of $\Gamma$ over the layer.
This exponential pressure increase implies, however, that such inversions 
are only  possible over a limited domain, since eventually the star 
must match an outer boundary condition of negligible pressure. 

Since more realistically the temperature should be expected to decline 
outward, such a pressure inversion would imply an even stronger 
outward increase in density. 
If the super-Eddington condition persists, then perhaps another 
switch to convective transport could again reduce the radiative flux, 
and bring the Eddington parameter back below unity.
But again this should become inefficient at a layer that cannot 
maintain an outflow against photon tiring.
This implies that any such outflow initiated from the region that 
convection becomes inefficient would necessarily stagnate at some 
finite radius.
One can imagine that the subsequent infall of material would likely 
form a complex spatial pattern, consisting of a mixture of both 
downdrafts and upflows, perhaps even resembling the 3-D cells of 
thermally driven convection.
Overall, we thus see that a star that exceeds the Eddington limit is 
likely to develop a complex spatial structure, whether due to local 
instability to convection, or to global instability of flow stagnation. 
Irrespectively, an inversion or efficient convection cannot offer a 
solution giving a photospheric luminosity which is super-Eddington.

\subsection{Lateral Instability of Thomson Atmosphere}

Dating back to work by \citet{Spiegel76,Spiegel77} there have been
speculations that an atmosphere supported by radiation pressure would 
likely exhibit instabilities not unlike Rayleigh-Taylor, associated with 
the support of a heavy fluid by a lighter one, leading to formation of 
``photon bubbles''. 
Recent quantitative stability analyses by \citet{ST99} and by \citet{Shaviv01ApJ} 
do lead to the conclusion that even a simple case of a pure ``Thomson atmosphere'' 
-- i.e., supported by Thomson scattering of radiation by free electron -- would be 
subject to intrinsic instabilities for development of lateral 
inhomogeneities.
The analysis by \citet{Shaviv01ApJ} suggests in particular that these
instabilities share many similar properties to the excitation of 
strange mode pulsations \citep[e.g.,][]{Glatzel94,Pap97}.
For example, they are favored when radiation pressure dominates over gas 
pressure. 
Both arise when the temperature perturbation term in the effective equation of 
state for the gas becomes non-local. 
In strange mode instabilities, the term arises because the temperature in the 
diffusion limit depends on the radial gradient of the opacity perturbations. 
In the lateral instability, the term depends on the lateral radiative flux which 
arises from non-radial  structure on a scale of the vertical scale height.

Note that when conditions of a pure Thomson atmosphere are alleviated, 
even more instabilities arise. 
There are of course the aforementioned strange mode instabilities,
which require a non-Thomson opacity. 
If magnetic fields are 
introduced, even more instabilities can play a role 
\citep{Arons92,Gammie98,Begelman02,Blaes03}. 
We stress however that the physical origin of the instabilities is not important 
to our discussion here. 
The essential point is merely that as atmospheres approach the Eddington limit, 
non-radial instabilities do exist to make the atmospheres inhomogeneous, 
while the typical length scale expected is that of the vertical scale height. 

\section{Super-Eddington Outflow Moderated by Porous Opacity}

\citet{Shaviv00,Shaviv01MNRAS} has applied these notions of a laterally 
inhomogeneous radiatively supported atmosphere to suggest a new
paradigm for how quasi-stationary wind outflows could be 
maintained from objects that formally exceed the Eddington limit. 
A key point regards the fact that, in a spatially inhomogeneous 
atmosphere, the radiative transport will selectively avoid regions of 
enhanced density in favor of relatively low-density, ``porous'' channels 
between them.
This stands in contrast to the usual picture of simple 1-D, gray-atmosphere 
models, wherein the requirements radiative equilibrium ensure that the 
radiative flux must be maintained independent of the medium's optical 
thickness.
In 2-D or 3-D porous media, even a gray opacity should lead to a flux 
avoidance of the most optically thick regions, much as in
frequency-dependent radiative transfer in 1-D atmosphere, wherein the 
flux avoids spectral lines or bound-free edges that represent 
localized spectral regions of non-gray enhancement in opacity.
The associated reduction in the effective opacity might therefore provide a
mechanism for the transition from an effectively sub-Eddington to 
super-Eddington condition, and perhaps thereby allow an 
appropriate regulation for steady wind mass loss of the outer layers.

\subsection{The Porosity Length and Effective Opacity of a Clumped Medium}

The radiative transport in such a complex, 3-D medium is likely to be extremely 
complicated, but to model the basic elements of this porosity 
effect, we can consider a simplified picture of a medium 
consisting of an ensemble of localized clumps or blobs.\footnote{
For comparison, \citet{Shaviv01MNRAS} considered two limiting cases, of either 
blobs in the optically thin limit or thick but vertically elongated, and a third
case of a two phase Markovian mixture. Our analysis here generalizes the first two 
cases, but is rather different than the statistical description in the third case.
Nonetheless, the comparison with the third case reaffirms that the mass loss scaling 
does not depend on the exact assumed geometry. 
}
Assuming for now that the blobs all have the same characteristic length 
$l$ and mass $m_{b}$, the characteristic blob density is
$\rho_{b} \approx m_{b}/l^{3}$, implying in a medium with 
opacity $\kappa$ a characteristic blob optical depth 
$\tau_{b} \approx \kappa \rho_{b} l$.
If the blobs are optically thick, $\tau_{b} \gg 1$, then their effective 
cross section to impingent radiation is just
$\sigma_{\eff} \approx l^{2}$.
But more generally, since for arbitrary optical thickness the fraction of 
impingent radiation attenuated by each blob should scale as 
$1-\exp(-\tau_{b})$, this effective cross section can be written as 
$\sigma_{\eff} \approx l^{2} [1-\exp(-\tau_{b})]$.
From this, we can accordingly define an {\it effective opacity} of the blobs as
\beq
\kappa_{\eff} \equiv { \sigma_{\eff} \over m_{b} } 
\approx {l^{2} \over m_{b}} [1-\exp(-\tau_{b})]
\approx \kappa \, { 1- \exp(-\tau_{b}) \over \tau_{b} } .
\label{keffdef}
\eeq
In the limit that the blobs are optically thin, $\tau_{b} \ll 1$,  
this effective opacity recovers the microscopic value, 
$\kappa_{\eff} \approx \kappa (1-\tau_{b}/2) \rightarrow \kappa$.
However for optically thick blobs, $\tau_{b} \gg 1$, the opacity is 
effectively {\it reduced by a factor $1/\tau_{b}$}, i.e.
$\kappa_{\eff} \rightarrow \kappa/\tau_{b} \approx l^{2}/m_{b}$.

Let us thus consider a medium that consists entirely of an ensemble of 
such blobs, with a characteristic separation scale $L \gg l$. 
Then the mean density of the medium is given by 
$\rho \approx m_{b}/L^{3} = (l/L)^{3} \rho_{b}$, 
and the blob optical thickness can be 
written as the ratio
\beq
\tau_{b} = {\rho \over \rho_{c}} \,
\label{taubdef}
\eeq
where the critical medium density at which the blobs have unit optical 
depth is given by
\beq
\rho_{c} 
\approx {l^{2} \over \kappa L^{3} }
\equiv { 1 \over \kappa h } \, ,
\label{rhocdef}
\eeq
with the latter equality defining a characteristic ``{\em porosity length}'' 
$h \equiv L^{3}/l^{2}$.

As developed further below, this porosity length turns out to be a key parameter 
for determining the nature and consequences of the porosity in a 
structured medium.
The above shows already that it can have a variety of interpretations:
\begin{itemize}
\item As defined here, it is the ratio of the volume per blob to the 
projected surface area of the blob, $h=L^{3}/l^{2}$.
\item Equivalently, it is the blob size divided by its volume 
filling factor, $h=l/(l/L)^{3}$.
\item As noted above, the inverse of the porosity length times the opacity 
defines  a critical mean density at which the blobs become optically thick, 
$\rho_{c} \equiv 1/(h\kappa)$.
\item For mean densities above this critical density $\rho \gg \rho_{c}$, it
represents the photon {\it mean-free-path} of the porous medium, 
$h = 1/ (\kappa_{\eff}\,  \rho )= 1/( \kappa \rho_{c})$, 
which consequently becomes {\it independent} of this mean density.
\end{itemize}

\subsection{Mass Loss Rate for Model with a Single Porosity Length}

Let us now apply this simple picture of a porous medium to model the 
porosity-moderated mass loss of a super-Eddington atmosphere.
The generic mass loss rate can be written as
\beqa
\nonumber
{\dot M} &=& 4 \pi \Rstar^{2} \rho_{\ast}   a_{\ast}
\\
\nonumber
&=& 
4 \pi \Rstar^{2} \rho_{\ast} a_{\ast} \, 
\left [ { 1 \over \rho_{c} \kappa h } \right ] \,
\left [ { H G \Mstar \over \Rstar^{2} a_{\ast}^{2} } \right ] \, 
\left [ {\kappa \Lstar \over \Gamma 4 \pi G \Mstar c } \right ]
\\
&=& 
\left ( { \rho_{\ast} \over \rho_{c} } \right ) \, 
\left ( {H \over h} \right ) \, 
\left ( {1 \over \Gamma } \right ) \,
{ \Lstar \over a_{\ast} c } 
\, ,
\label {mdgens}
\eeqa
where the basic definition in the first line is multiplied in the 
second line by a series of unity factors (in square brackets), 
which are defined in terms of 
the porosity length $h$, 
the critical density  $\rho_{c}$ ($=1/\kappa h$), 
the gravitational scale height $H$ ($=a_{\ast} \Rstar^{2}/G\Mstar$), 
and 
the Eddington parameter $\Gamma$ ($=\kappa \Lstar/4 \pi G\Mstar c$).
After several cancellations, the result in the third line 
yields three {\em dimensionless} scale factors (in parenthesis) that 
multiply the {\em dimensional} scaling set by
the luminosity divided by the product of the speed of light times sound speed.
For future reference, we refer to this last factor as the ``basal'' mass 
loss rate, with notation
\beq
{\dot M}_{\ast} \equiv { \Lstar \over a_{\ast} c }  \, .
\label{mdbasal}
\eeq

So far this is just a simple recasting of the generic mass loss 
scaling, but it now allows us to apply properties of a specific 
porosity model to derive associated physical scalings of mass loss 
from a porosity-moderated, super-Eddington medium.

Let us then first define a scaling for the porosity length $h$.
For this,  consider arguments analogous to those
traditionally given for representing convective energy transport in terms 
of a characteristic ``mixing length'', which is generally assumed to scale in 
proportion to the gravitational scale height $H$.
Specifically, let us make an analogous {\it ansatz} that the 
porosity length should likewise scale with this pressure scale height.
We thereby define a dimensionless ``{\em porosity-length parameter}'',
\beq
\eta \equiv { h \over H } \, ,
\label{poransatz}
\eeq
which is analogous to the mixing-length parameter, $\alpha =l_{mix}/H$, 
traditionally assumed for treating convective energy transport.

Note that this scaling relates the porosity length to the 
pure gas pressure scale height, ignoring any modification due to 
radiation pressure.
A basic justification for this lies in the expectation that
the effective Eddington parameter will be substantially reduced in the
porous medium, so that in the final self-consistent state the medium again 
becomes stratified on a scale proportional to $H$.
More generally, any effect of radiation pressure in setting this 
equilibrium can be accounted for by taking this porosity-length parameter 
to be a function of the Eddington parameter, $\eta[\Gamma]$.
However, for simplicity, we typically take $\eta$ to be a fixed 
independent parameter in the analysis here.

Let us next derive the scaling for the sonic density ratio,
$\rho_{\ast}/\rho_{c}$.
Given a value of the Eddington parameter $\Gamma > 1$ defined from 
a microscopic continuum opacity $\kappa$, the reduced, effective 
Eddington parameter for material with arbitrary density $\rho$ is 
simply given from application of 
eqns.~(\ref{keffdef}) and (\ref{taubdef}),
\beq
\Gamma_{\eff}   = {\kappa_{\eff} \over \kappa} \Gamma \approx 
{\rho_{c} \over \rho} \Gamma 
\left ( 1-e^{-\rho/\rho_{c}} \right ) 
\label{geffdef}
\, .
\eeq
The reduction at large densities now allows a base hydrostatic 
region where $\Gamma_{\eff} < 1$. 
The transonic wind is then initiated at the point where 
$\Gamma_{\eff} = 1$, which when applied to eqn.~(\ref{geffdef}) 
defines an implicit relation for the sonic point density $\rho_{\ast}$.
An {\it explicit} solution 
is given by
\beq
{\rho_{\ast} \over \rho_{c}} = \Gamma + {\rm W_{0}}
\left [ -\Gamma \exp(-\Gamma) \right ] \, 
\approx \Gamma - 1/\Gamma
\label{dsexsoln}
\eeq
where ${\rm W_{0}}$ represents the principal branch of the
{\em Product-Log} (a.k.a.\ Lambert) function\footnote{
{\tt http://mathworld.wolfram.com/LambertW-Function.html}
} \citep{JHC96},
and the latter, much simpler, approximate form is valid within 6\% of the exact 
solution for all the $\Gamma > 1$ that are of interest.

Application of the approximate form from (\ref{dsexsoln}) into (\ref{mdgens}) 
yields the mass loss scaling,
\beqa
{\dot M_{por}} 
&\approx& 
\left ( 1 - {1 \over \Gamma^{2}} \right ) \, 
{ \Lstar \over \eta a_{\ast} c}  
\label{Mdotpor1}
\\
&=& 
\left ( 
{ \Gamma + 1 \over \eta \Gamma} 
\right )
{ \Lstar- L_{Edd} \over a_{\ast} c}  
\, 
\label{Mdotpor2}
\\
&=&
~~~ \wf [\Gamma] ~~~
{ \Lstar- L_{Edd} \over a_{\ast} c}  
\, .
\label{Mdotpor3}
\eeqa
The second equality recasts this scaling in terms of a difference of 
the luminosity from 
the Eddington luminosity,
$L_{Edd} \equiv 4 \pi G\Mstar c/\kappa$.
This is quite similar to the scaling derived by \citet{Shaviv01MNRAS}, 
with the final equality introducing his ``wind function'' 
(cf.\  his eqn.~30).
In the present formalism this is seen to have the approximate scaling 
%
$\wf(\Gamma) \approx (\Gamma+1)/\Gamma \eta$,
which hence varies 
from 
$\wf \approx 2/\eta$ for the marginally super-Eddington case $\Gamma-1 \ll 1$ 
to 
$\wf \approx 1/\eta$ for the strong super-Eddington limit $\Gamma \gg 1$.

In terms of the luminosity-proportional form (\ref{Mdotpor1}), the dimensional
values of the mass loss rate for this  single-porosity-length model scale with 
the basal rate defined in eqn.~(\ref{mdbasal})  
\beqa
 { \eta {\dot M_{por}} \over 1-1/\Gamma^{2}} 
\approx
{\dot M}_{\ast} 
&\equiv&
{\Lstar \over a_{\ast} c} 
\nonumber
\\
&=&
0.001 \, {\Msun \over {\rm yr}} 
\, {L_{6} \over a_{20} }
\, ,
\label{Mdotgm1}
\eeqa
where 
$a_{20} \equiv a_{\ast}/20$~km/s
and $L_{6} \equiv \Lstar/10^{6} \Lsun$.
Dividing this basal mass loss rate by the tiring-limited value in eqn.~
(\ref{mdtir}) gives for the associated ``tiring number''
\beq
m_{\ast} \equiv { {\dot M_{\ast}} \over {\dot M}_{tir}} 
= {v_{esc}^{2} \over 2 a_{\ast} c} 
 = {  0.032  \over a_{20} } ~ {\Mstar/\Rstar \over \Msun/\Rsun } 
\, .
\label{mbasal}
\eeq
We thus see that, under the canonical assumption that $\eta \approx 1$
(i.e., $h \approx H$),
this single-porosity-length model yields a maximum mass loss 
rate that is a few percent of the photon tiring limit.

\subsection{Two-Component Model of Clumps in a Smooth Background Medium}

As a next step in developing this phenomenological model for porosity, 
consider the more realistic case in which these localized blobs are 
embedded within a smooth, unclumped component.
Let us again assume the clumped component consists of many individual
blobs with mass $m_b$ confined to a small size $l \ll L$ compared 
to the interblob spacing $L$, giving them
a density $\rho_b = m_b/l^3$ that is much larger than the interblob
density $\rho_i = m_i/L^3$.
The density averaged over blob and interblob medium is 
\beq
\bar{\rho} ={ m \over L^3} = \rho_i + \rho_b (l/L)^3 ,
\label{1.1}
\eeq
where the total mass $m \equiv m_i+m_b$.

Again taking the medium to have a fixed
microscopic opacity $\kappa$,
the individual blobs thus now have an optical thickness
\beq
\tau_b = \kappa \rho_b l = \kappa m_b/l^2 = 
\kappa {\bar \rho} f L^3/l^2 = \kappa {\bar \rho} f h  ,
\label{1.2}
\eeq
where $f \equiv m_b/m$ is the blob mass fraction, and 
$h \equiv L^3/l^2$ is again the characteristic porosity length.
As long as the blobs are optically thin, $\tau_b \ll 1$, both the blobs and 
the interblob medium have this same effective mass-absorption coefficient, 
namely $\kappa$.
But if the blobs become optically thick, $\tau_b \ge 1 $, 
then, compared to a uniform medium, the effective opacity of this
clumped medium is now reduced by a factor
\beqa
k &\equiv& {\kappa_{\eff} \over \kappa} 
\nonumber
\\
&=& 
{ \kappa m_i + \kappa_b m_b \over \kappa m} 
\nonumber
\\
&=& 
\, 1-f \,  + \, f \,  { 1-e^{-\tau_b} \over \tau_b  } 
\nonumber
\\
&\equiv& 
\, k_{min} + k_{b}
\, .
\label{1.4}
\eeqa
Here $k_{min} \equiv 1-f$ represesent a minimum value of this 
effective opacity ratio associated with the interblob medium, 
while $k_{b}$  represents the reduced opacity of the blob component.

It is again instructive to consider a few limiting cases.
First, for optically thin blobs $\tau_b \ll 1$, we do indeed recover
that there is no effective change
\beq
k \approx 1 - {f \tau_b \over 2} +{\cal O}(\tau_b^2) \rightarrow 1 
~ ; ~ \tau_b \ll 1  .
\label{1.5}
\eeq
In the opposite extreme that the medium consists almost entirely
of optically thick blobs, with very little interblob material, 
$1-f \ll 1/\tau_b$, we again find that effective absorption is reduced by
a factor given by the inverse of the individual blob optical thickness,
\beq
k \rightarrow {1 \over \tau_b} ~ ; ~
 1/(1-f) \gg \tau_b \gg 1 ,
\label{1.12}
\eeq
which in principal can imply quite a substantial reduction in absorption.
In the somewhat intermediate case of optically thick blobs with an interblob 
medium that has a non-negligible mass fraction $1-f \gg 1/\tau_b$, the 
absorption is dominated by this interblob material, yielding
\beq
k \rightarrow k_{min} ~; ~ \tau_b \gg 1/(1-f) > 1.
\label{1.13}
\eeq
This last case represents an important component of added realism for 
the model, since it implies that, unlike the pure-clumped case 
considered above, the porosity reduction does not continue to scale in
proportion to the inverse density to arbitrarily small values, but rather 
saturates to a minimum or ``floor'', $k_{min} = 1-f$.
More realistically, instead of identical clumps, it seems likely that 
the structured component may be better modeled as consisting of a broad
{\it distribution} in clump properties, as is developed in the next 
section.

The general picture then is that, in such a porous medium, the 
radiative flux, and thus the radiative acceleration, should be weighted 
towards the lower-density regions, which, in principle can thus have an 
even greater (super-Eddington) outward driving.
However, a further fundamental assumption in the analysis in this 
paper is that overall medium can  still be described in terms of a 
{\it single-fluid} model.
One justification lies in the relatively high overall densities 
in the regions of flow initiation -- which thus implies a strong gas 
collisional coupling to share momentum between lower and higher 
density components.
Moreover, since many of the models for structure formation involve 
travelling wave solutions, the actual material can even alternate in 
time between dense and rarefied regions, thus implying a further overall 
averaging in radiative momentum addition.
With this overall assumption of strong effective momentum coupling to 
maintain a  single-fluid medium, we thus leave to future studies the issue 
of possible dynamical differentiation between regions of lower and higher 
density.

\section{Super-Eddington Wind Moderated by a Power-law Porosity}

\subsection{A Power-Law-Porosity Ansatz}

In considering options for further development of this basic formalism, 
let us first note here an interesting similarity in how both continuum 
porosity and line opacity lead to a reduced effectiveness in
absorption as a function of the relevant optical thickness 
parameter, viz. 
$\tau_{b} = \kappa \rho h$ vs.~
$\tau_{l}\equiv q \kappa_{e} \rho c/(dv/dr)$
(see \S 2.2).
Namely, in both cases, the correction factor takes the form
$(1-e^{-\tau})/\tau$; cf.\  eqns.~(\ref{gsingline}) and (\ref{keffdef}).

Now, as noted in \S 2.2, in the theory of line-driven winds 
a crucial extension over early, single-line analyses
\citep{LS70} was the introduction by CAK of a formalism 
for treating the cumulative effect of many lines.

Likewise, recognizing the unrealistic nature of the above simple 
picture of a clumped medium in which all the blobs have identical 
properties, let us next consider a model in which the structured
component consists of an {\it ensemble} of individual clumps with 
a {\it range} of optical depths $\tau_i$. 
Compared to a smooth, unclumped medium, the effective opacity of the 
clump component should then be reduced by a factor 
\beq
k_{b} \equiv {\kappa_{\eff} \over \kappa} = 
\sum_i f_i { 1 - e^{-\tau_i} \over \tau_i} 
\approx \int_0^\infty d\tau \, {df \over d\tau} \, 
\left ( { 1 - e^{-\tau} \over \tau} \right ) ,
\label{3.1}
\eeq
where the latter approximation assumes the summation can be approximated by an 
integral over a suitably defined clump distribution in optical depth, $df/d\tau$.

In direct analogy with the CAK power-law formalism for line-strength,
let us specifically make the {\it ansatz} that this distribution can be described by 
an exponentially truncated power law 
(see \S\S 2.2 and 2.3, and eqn.~\ref{dndqcak}),
\beq
\tau {df \over d\tau} = { 1 \over  \Gamma [\alpha_{p}]} \,
\left ( { \tau \over \tau_o} \right )^{\alpha_{p}} e^{-\tau/\tau_o} ,
\label{3.2}
\eeq
where $\tau_o$ now represents the optical depth of the {\it strongest} clump, 
and $\alpha_{p} > 0$ is the power-index\footnote{
We assume that $\alpha_p>0$ because otherwise the contribution from
weak structure causes the overall number normalization integral (\ref{3.3}) 
to diverge, unless one introduced a further free parameter to cut off 
the distribution at some minimum porosity length.
The subscript ``p'' stands for porosity, and is added to distinguish 
this exponent from the usual CAK line-opacity power-index $\alpha$. 
Finally, note that $\Gamma[\alpha_{p}]$ represents here the Gamma function, 
and not the Eddington parameter.}.
This assumed distribution obeys the normalization 
\beq
\int_0^\infty d\tau \, {df \over d\tau} \equiv 1 .
\label{3.3}
\eeq

Applying eqn.~(\ref{3.2}) in eqn.~(\ref{3.1}), we find the opacity 
reduction of the clumped medium
to be given by (cf.\  eqn.~\ref{gamcak}), for $\alpha_{p} \ne 1$,
\beqa
k_{b} [\tau_o ] &=& { (1+\tau_o)^{1-\alpha_{p}} - 1 \over (1-\alpha_{p}) \tau_o } 
\nonumber
\\
&=&
{ 
\left [ 
(1+1/\tau_o)^{1-\alpha_{p}} - (1/\tau_o)^{1-\alpha_{p}} 
\right ]
\over (1-\alpha_{p}) \tau_o^{\alpha_{p}} 
} \, 
,
\label{3.4}
\eeqa
or for the special case $\alpha_{p}=1$,
\beq
k_{b} [\tau_o ] = { \ln (1+\tau_{o}) \over \tau_{o} } 
.
\label{3.4.5}
\eeq
For fixed clump characteristics, the optical depth $\tau_o$ 
again (cf.\  eqn.~\ref{taubdef})
scales with the density $\rho$,
\beq
\tau_o \equiv { \rho \over \rho_o }
\label{3.7}
\eeq 
where the critical density at which this 
strongest blob has unit optical depth is (cf.\  eqn.~\ref{rhocdef})
\beq
\rho_{o} = { 1 \over \kappa h_{o} } ,
\eeq
with $h_{o}$ the associated porosity length.
This model of a medium with a power-law porosity is now characterized by 
two parameters, the power index $\alpha_{p}$,  and the 
porosity length $h_{o}$ of the strongest blob.

In the limit that even the strongest clump is optically thin, 
$\tau_o \ll 1 $, we recover from eqns.~(\ref{3.4}) and (\ref{3.4.5}) 
that there is only a small reduction in opacity,
\beq
k_{b} [\tau_o] \approx 
1- \alpha_{p} \tau_o/2
~~ ; ~~ \tau_o \ll 1 .
\label{3.6}
\eeq

In the opposite limit that the strongest clump is optically thick, 
$\tau_o \gg 1 $, the asymptotic scaling depends on whether 
$\alpha_{p}$ is larger or smaller than one.
For $\alpha_{p} <1$, the square bracket term in (\ref{3.4}) approaches unity
in this thick limit, and we obtain for the opacity reduction,
\beq
k_{b} [\tau_o]\approx  
{ 1 \over (1-\alpha_{p}) \tau_o^{\alpha_{p}} } 
~~ ; ~~ \tau_o \gg 1~\&~~\alpha_p <1 \, .
\label{3.5}
\eeq
For $\alpha > 1$, we find
\beq
k_{b} [\tau_o] \approx  
{ 1 \over (\alpha_{p}-1) \tau_o } ~~ ; ~~ \tau_o \gg 1 ~\&~~\alpha_p >1.
\label{3.5.1}
\eeq
As seen from eqn.~(\ref{3.4.5}), the special case $\alpha_{p}=1$ also scales as 
$k_{b}(\tau_{o}) \propto 1/\tau_{o}$ in this optically thick limit, but with a 
proportionality factor that increases logarithmically with optical depth, 
i.e., $\ln(1+\tau_{o})$.

We thus see that the effective opacity of the structured component 
can become arbitrarily small at large densities, 
$\rho / \rho_{o} = \tau_{o} \gg 1 $.
However, any ``interclump'', {\em smooth} compoent should still 
contribute a constant opacity, independent of density, so that the overall 
effective opacity is given by
\beq
k [\tau_{o} ] = k_{min} + 
k_{b} [\tau_{o}] 
\, ,
\label{ktotpow}
\eeq
which thus has a ``floor'' or minimum effective opacity, $k_{min}$, 
set the mass fraction of any smooth, interclump component.

\subsection{Mass Loss Rate for the Power-Law Porosity Model}

Let us now apply this power-law porosity model toward computing the 
mass loss properties from a super-Eddington surface layer.
In analogy with eqn.~(\ref{mdgens}), we first write the generic mass loss rate in
a form that is scaled by the porosity-length $h_{o}$ and the 
associated critical density $\rho_{o} = 1/\kappa h_{o}$, 
\beq
{\dot M} = 
\left ( { \rho_{\ast} \over \rho_{o} } \right ) \, 
\left ( {H \over h_{o}} \right ) \, 
\left ( {1 \over \Gamma } \right ) \,
{ \Lstar \over a_{\ast} c } 
\, .
\label{mdgenspow}
\eeq
The sonic density $\rho_{\ast}$ is now found through application of eqn.~
(\ref{3.4}) or (\ref{3.4.5}) to the sonic point condition,
\beq
\Gamma \, k[\rho_{\ast}/\rho_{o}] \, = \, 1 
\, ,
\label{3.11}
\eeq
where now we must also require the mass fraction of interclump
medium $k_{min} < 1/\Gamma$, so that eqn.~
(\ref{3.11}) has a solution. 
In the following analyses, we assume this is always the case, 
regardless of how large $\Gamma$ becomes.
In effect, we thus henceforth effectively take
\beq
k[\rho/\rho_{o}] \approx k_{b} [\rho/\rho_{o}] ~~ ; ~~ \rho \le \rho_{\ast}
\, ,
\eeq
which applies from the sonic point outward, i.e., through any resulting
wind outflow.

In such a medium in which the unclumped Eddington parameter $\Gamma$ is 
sufficiently above unity, this sonic point is located where 
the density is high enough that the strongest clump is quite 
optically thick, $\tau_{o} \gg 1$.
Under this circumstance, the structured component opacity reduction
given by the full form (\ref{3.4}) can be reduced to the simple power-law forms 
(\ref{3.5}) and (\ref{3.5.1}), 
thereby allowing trivial solution of the critical condition (\ref{3.11})
\beqa
{ \rho_{\ast} \over \rho_{o} } 
&\approx&
\left [ \Gamma \over 1 - \alpha_{p} \right ]^{1/\alpha_{p}}  
~~ ; ~~ \Gamma \gg 1 ~ \& ~ \alpha_{p} < 1
\, ,
\label{tsgbig}
\\
& \approx & 
\left [ { \Gamma \over \alpha_{p} - 1 } \right ]
~~~~~~~ ; ~~ \Gamma \gg 1 ~ \& ~\alpha_{p} > 1
\, .
\label{tsgbig2}
\eeqa
In the opposite limit for which the Eddington parameter is only marginally 
above unity, we find through application of eqn.~(\ref{3.6}) in eqn.~
(\ref{3.11}),
\beq
{ \rho_{\ast} \over \rho_{o} } 
\approx { 2 (\Gamma - 1) \over \alpha_{p}} 
~~ ; ~~ \Gamma - 1 \ll 1 ~ .
\label{tsgapp1}
\eeq
For the special case $\alpha_{p}=1/2$, a general solution to eqn.~
(\ref{3.11}) is
\beq
{ \rho_{\ast} \over \rho_{o} }  =  4 \Gamma (\Gamma - 1)  
~~ ; ~~ \alpha_{p}  = 1/2 ~ .
\label{tspowh}
\eeq
For the special case $\alpha_{p}=1$, application of eqn.~
(\ref{3.4.5}) in (\ref{3.11}) also yields a general, explicit solution,
\beq
{ \rho_{\ast} \over \rho_{o} }  =  
-1 - \Gamma W_{-1} [-\exp(-1/\Gamma)/\Gamma] 
\approx 2\Gamma \log(\Gamma)
~~ ; ~~ \alpha_{p}  = 1 ~ ,
\label{tspow1}
\eeq
where $W_{-1}$ is now the lower branch of the Product-Log (Lambert) 
function \citep{JHC96}, and 
the latter approximation is roughly applicable for moderate $\Gamma$,
e.g., remaining
within ca.\ 25\% of the correct solution for $\Gamma \le 10$.

Next, let us again introduce a {\it porosity-length parameter}, 
$\eta_{o} \equiv h_{o}/H$, relating now to the {\it strongest} clump 
(cf.\  eqn.~\ref{poransatz}).
Applying this, plus eqns.~(\ref{tsgbig})-(\ref{tspow1}), into eqn.~
(\ref{mdgenspow}), we find the mass-loss rate for the power-law 
porosity model has the scalings
\beqa
{\dot M_{pow}} 
&\approx&
{1 \over \Gamma}
\left [ \Gamma \over 1 - \alpha_{p} \right ]^{1/\alpha_{p}} 
{ \Lstar \over \eta_{o} a_{\ast} c }
~ ;  ~~ \Gamma \gg 1  ~ \& ~ \alpha_{p} < 1  ~~~~~~~~~
\label{Mdpowpora}
\\
&\approx&
~~~~
{1 \over \alpha_{p} - 1 }
~~~~~~~~ { \Lstar \over \eta_{o} a_{\ast} c }
~ ; ~~ \Gamma \gg 1  ~ \& ~ \alpha_{p} > 1 \
\label{Mdpowporb}
\\
&\approx&
~~~
{2 (\Gamma - 1) \over \alpha_{p}} 
~~~~~~ { \Lstar \over \eta_{o} a_{\ast} c }
~~ ; ~~ \Gamma - 1 \ll 1 
\label{Mdpowporc}
\\
&\approx&
~~~~
2 \log (\Gamma) 
~~~~~~~ { \Lstar \over \eta_{o} a_{\ast} c }
~~ ; ~~ \alpha_{p}  = 1 
\label{Mdpowpord}
\\
&=&
~~~~
4(\Gamma - 1) 
~~~~~~ { \Lstar \over \eta_{o} a_{\ast} c }
~~ ; ~~ \alpha_{p}  = 1/2 ~ .
\label{Mdpowpore}
\eeqa

It is of interest to compare these results with those given in
eqn.~(\ref{Mdotpor1}) for the single-length porosity model of \S 5.2.
For $\alpha_{p} > 1$, eqns.~(\ref{Mdpowporb}) and (\ref{Mdpowporc})
show that the scaling is similar to the single-length model in both 
the strongly ($\Gamma \gg 1$) and weakly ($\Gamma-1 \ll 1$) 
super-Eddington limits.
As noted above, this stems from the fact that in such $\alpha_{p}>1$ models 
the opacity reduction is dominated by structure with the largest 
porosity length.

However, for $\alpha_{p} \le 1$, comparison of eqn.~(\ref{Mdpowpora}) 
and (\ref{Mdotpor1}) shows there can  be a substantially
larger mass loss in the strongly super-Eddington limit of the 
power-law model.
In the single-length model, the 
opacity in dense layers 
is reduced by the inverse of the density, but in the $\alpha_{p} < 1$ 
power-law models the reduction is weaker, 
scaling as $1/\rho^{\alpha_{p}}$.
For a large overall Eddington parameter $\Gamma \gg 1 $, the 
sonic-point reduction of the effective Eddington to unity  
occurs at a deeper, denser layer, implying a larger mass loss rate.
The net result is to increase the sensitivity of the mass loss rate 
to the Eddington parameter $\Gamma$.
For example, for $\alpha_{p}=1/2$, the overall scaling is in proportion to 
$4 (\Gamma -1)$ instead of $1-1/\Gamma^{2}$.

Finally, let us consider this power-law-porosity mass-loss relative to the 
tiring limit discussed in \S 3.
From the discussion in \S 5.2 for the single-porosity length model, 
eqn.~(\ref{mbasal}) already gives the scaling of the tiring number 
associated with the {\it basal} mass loss rate, 
$\Mdot_{\ast}=\Lstar/a_{\ast}c$.
In the power-law models, the full tiring number is just increased by 
the additional dimensionless factors from 
eqns.~(\ref{Mdpowpora})-(\ref{Mdpowpore}).
For example, for the analytic case $\alpha_{p}=1/2$, we find the 
tiring number scales as
\beqa
m_{pow} &\equiv&{ {\dot M_{pow}} \over {\dot M}_{tir}} 
\nonumber
\\
&\approx& 4 (\Gamma - 1 ) { m_{\ast} \over \eta_{o} }
\nonumber
\\
&=&  0.13 \, 
{\Gamma - 1 \over \eta_{o} a_{20} } \, {\Mstar/\Rstar \over \Msun/\Rsun }
~~ ; ~~ \alpha_{p} = 1/2
\, .
\label{mpow}
\eeqa
We therefore see that, under the canonical assumption that $\eta_{o} \approx 1$
(i.e., $h_{o} \approx H$) and $a_{20} \approx 1$,
a porosity model with power index $\alpha_{p}=1/2$ and 
a  moderately large Eddington parameter
can readily approach the tiring limit, e.g., $m_{pow} \approx 
0.52$ for $\Gamma = 5$.

Eqn.~(\ref{mpow}) suggests that while photon tiring should have marginal 
influence for only mildly super-Eddington models, i.e., $\Gamma \ltwig 2$,
it should limit the effective mass loss in strongly 
super-Eddington cases, i.e., $\Gamma > 3$.
We  examine this issue further in conjunction with determination of the 
associated wind velocity law (\S 6.4).

\subsection{Wind Velocity Laws for Power-Law Porosity Models without 
Tiring}

In addition to mass loss rate, the power-law porosity formalism can also 
be used to model the outward wind acceleration and resulting velocity 
law.
Let us first examine this under the assumption that the mass loss rate 
is small enough that photon tiring can be neglected.
Ignoring as before gas pressure effects, 
the wind acceleration can now be described by the dimensionless equation of
motion,
\beq
w' (x) = \Gamma k[\tau_o(x)] - 1 .
\label{3.9}
\eeq
As above, $w'$ is the wind acceleration in units of the local gravitational 
acceleration, and $x$ is a spatial coordinate defined by 
$x \equiv 1-\Rstar/r$, where $r$ is the local radius and $\Rstar$ is a 
characteristic wind sonic-point radius at which the 
RHS
crosses zero, i.e., $w'(0) = \Gamma k[\tau_o(0)] -1 = 0$.
Applying 
the opacity reduction eqn.~(\ref{3.4}), 
the equation of motion (\ref{3.9}) 
becomes an ordinary, first-order differential equation that can be integrated 
using standard numerical techniques.
The free parameters 
in this integration are 
the power-law index $\alpha_{p}$, the Eddington parameter $\Gamma$, and 
the initial flow energy $w(0)=w_{\ast} $.

Figure 2 shows results for the ratio of wind terminal speed to escape 
speed $v_{\infty}/v_{esc}$ ($\equiv \sqrt{w[1]}$), plotted vs.~
Eddington parameter $\Gamma$ for various $\alpha_{p}$ and $w_{\ast}$.
The terminal speeds are typically of order the escape speed, with the larger 
values for larger $\Gamma$, larger $\alpha_{p}$, and lower $w_{\ast}$.

\begin{figure}
\plotone{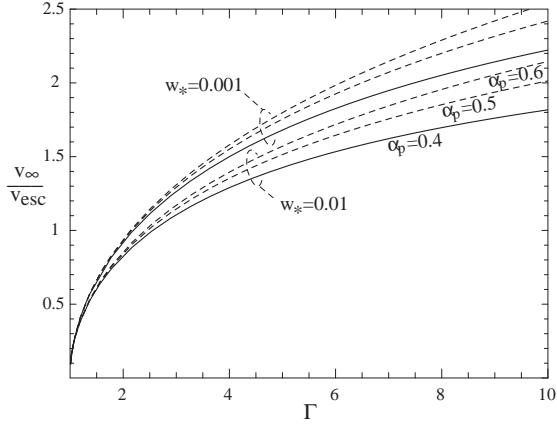}
\caption{
Ratio of terminal speed to escape speed, $v_{\infty}/v_{esc}$  
[$\equiv \sqrt{w(1)}$~], plotted as a function of Eddington 
parameter $\Gamma$, for selected values of $\alpha_{p}=$~0.4, 0.5, and 0.6 
(lower, middle, and upper curves of each triad), and 
for initial energy $w_{\ast}=0.01$ (lower triad) and 
$w_{\ast} = 0.001$ (upper triad).
The results can be roughly fit by
$v_{\infty}/v_{esc} \approx [\log(1/w_{\ast}) (\Gamma-1)/3]^{(1+\alpha_{p})/4}$.
}
\end{figure}

We also generally find that the height variation of velocity can be 
reasonably well fit by a modified ``beta'' velocity law of the form
\beqa
v(r) &=&
\sqrt{ a^2 + (v_\infty^2 - a^2)  \left ( 1 - {\Rstar \over r} \right 
)^{2 \beta} }
\nonumber
\\
&\approx&
v_\infty \left ( 1 - {\Rstar \over r} \right )^{\beta} \, ,
\label{3.13}
\eeqa
with however the velocity power index $\beta$ dependent on the 
parameters $\Gamma$, $\alpha_{p}$, and $w_{\ast}$.
To characterize this dependence, we define an effective index 
\beq
\beta_{\eff} \equiv 0.5 \log{
\left [
{ w(1)-w_{\ast} \over 
w(0.1)-w_{\ast} } 
\right] } \, ,
\label{beffdef}
\eeq
which we find gives a reasonable fit to the numerically integrated 
velocity variation.
Fig.~3 plots $\beta_{\eff}$ vs.~$\Gamma$ for various $\alpha_{p}$ and 
$w_{\ast}$.
The values are typically within 25\% of unity, with larger $\beta_{\eff}$ 
occurring for larger $\Gamma$, larger $w_{\ast}$, and smaller $\alpha_{p}$.

\begin{figure}
\plotone{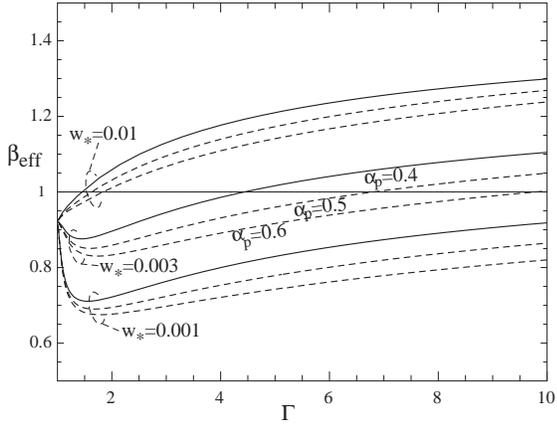}
\caption{
Effective velocity law index $\beta_{\eff}$ 
($\equiv 0.5 \log[(w(1)-w_{\ast})/(w(0.1)-w_{\ast})]$)
plotted as a function of Eddington parameter $\Gamma$, for 
selected values of $\alpha_{p}=$~0.4, 0.5, and 0.6 (lower, middle, and 
upper curves of each triad), and for initial energy 
$w_{\ast} = 0.01 $ (upper triad), 
$w_{\ast} = 0.003$ (middle triad), and 
$w_{\ast} = 0.001$ (lower triad).
}
\end{figure}

By comparison, for line-driven winds typical terminal speeds are 
2-3 times the escape speed, with a velocity-law power index $\beta 
\approx 0.8-1$.
Figs. 2 and 3 show similar results for the power-law porosity model of 
continuum driving, with however a somewhat wider range that depends on 
the parameters $\Gamma$, $\alpha_{p}$, and $w_{\ast}$.

\subsection{Photon Tiring and Wind Stagnation for Strongly 
Super-Eddington Porosity Models}

As noted in \S 6.2, the mass loss rates expected for power-law 
porosity models with moderately large Eddington parameters 
($\Gamma \gtwig 3-4$) approach the photon tiring limit.
Let us now derive velocity solutions for porosity models that 
take into account photon tiring.
From eqns.~(\ref{eomtir}) and (\ref{3.4}) the equation of motion now
takes the form
\beq
w'(x)
= -1 +  k[\tau_{o}(x)] \, \Gamma \, [1 - m (w+x)] .
\label{eomportir}
\eeq
For a given tiring number $m$, this can again be integrated numerically
from the initial value $w_{\ast}$, given also the model parameters 
$\Gamma$ and $\alpha_{p}$,

\begin{figure}
\plotone{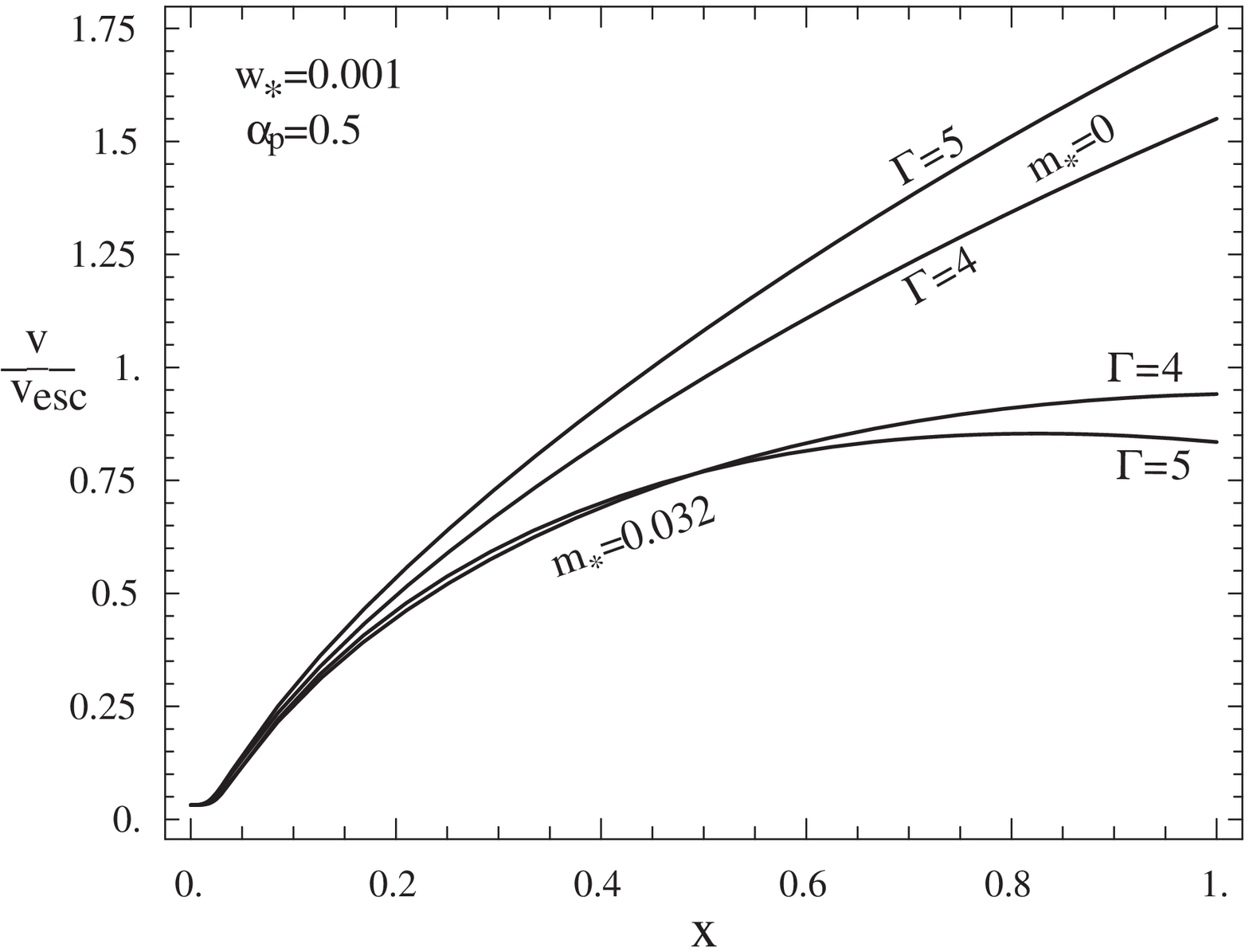}
\plotone{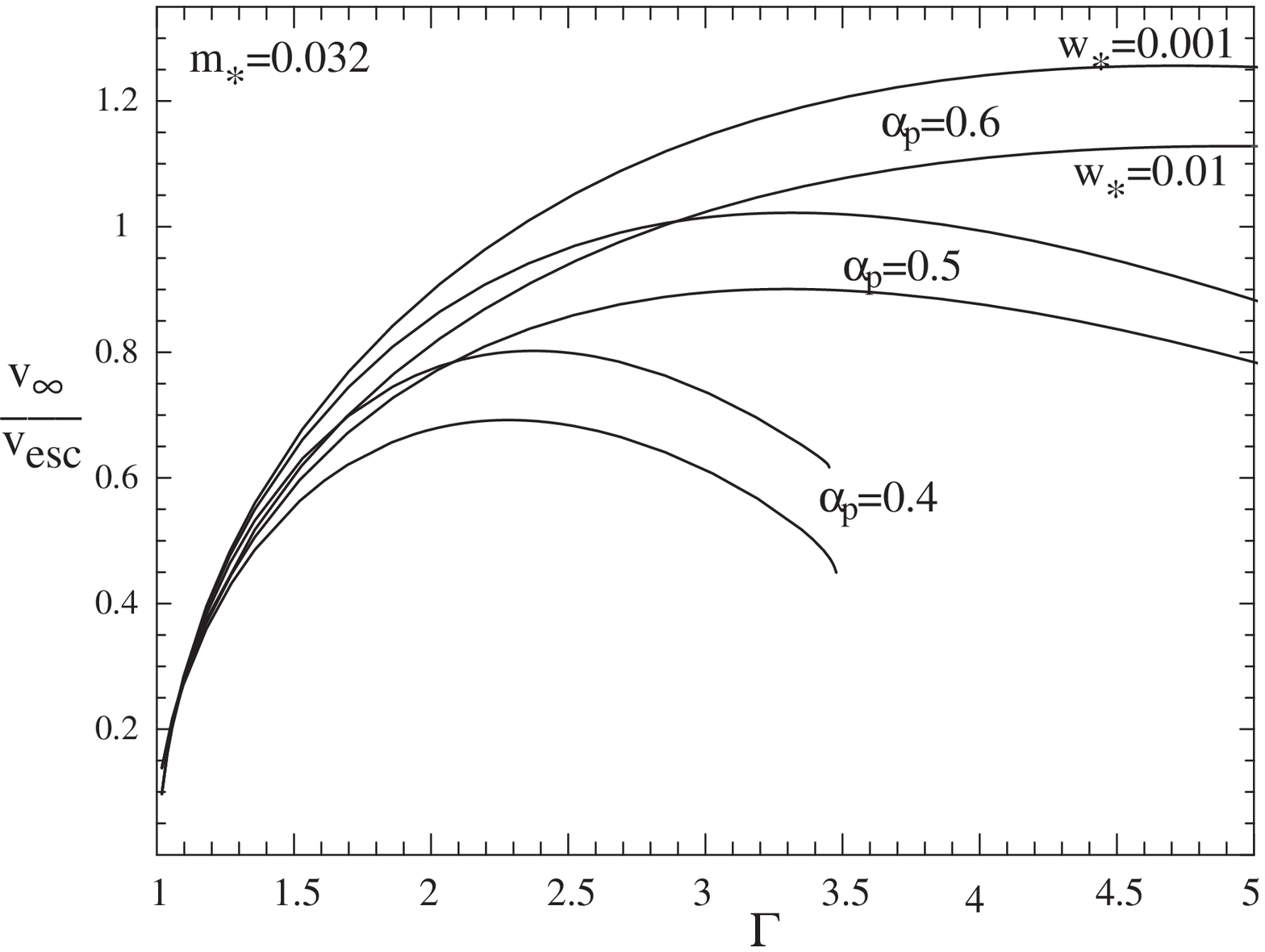}
\caption{
a.) Velocity over escape velocity plotted vs.~scaled inverse radius $x$, 
for Eddington parameters $\Gamma$=4 and 5, for models ignoring photon 
tiring ($m_{\ast}=0$) and with basal tiring parameter 
$m_{\ast} \equiv v_{esc}^{2}/2a_{\ast}c = 0.032$.
b.) Terminal speed over escape speed, plotted vs.~Eddington 
parameter $\Gamma$, for models with basal tiring parameter $m_{\ast} = 
0.032$, 
with various power-indexes $\alpha_{p}$ and initial energies $w_{\ast}$.
}
\end{figure}

Using eqns.~(\ref{mbasal}) and (\ref{mdgenspow}), we can write the tiring 
number as
\beq
m ={\rho_{\ast}   m_{\ast} \over \rho_{o} \eta_{o} \Gamma }
\label{mtdef}
\eeq
where $m_{\ast} \equiv v_{esc}^{2}/2a_{\ast}c = a_{\ast}/2w_{\ast} \, c$, 
and $\rho_{\ast}/\rho_{o}$ is a set by $\alpha$ and $\Gamma$ through solution 
of eqn.~(\ref{3.11}). 
For the typical parameters $\alpha_{p} = 1/2$ and $w_{\ast}=0.001$,
figure 4a then plots the velocity over escape velocity, $v/v_{esc}$, vs.~
scaled inverse radius $x$, comparing results without tiring ($m_{\ast}=0$)
and with the  fiducial value of the basal tiring number $m_{\ast}=0.032$ 
(eqn.~\ref{mbasal}).
Note that without tiring the velocity is higher for the higher 
Eddington parameter, 
but with tiring the velocity 
is reduced, with the stronger reduction 
now making the velocity {\it lower} for $\Gamma=$~5 than for 
$\Gamma=$~4.

For this same basal tiring number $m_{\ast}=0.032$, Fig.~4b shows the 
terminal speed over escape speed, $v_{\infty}/v_{esc}$, plotted vs.~
the Eddington parameter $\Gamma$, 
for selected parameters $\alpha_{p}$ and $w_{\ast}$.
Again the tiring reduces the speed, but note that for the lower power 
index, $\alpha_{p}=0.4$ with moderately 
large $\Gamma > 3.5$, the wind can no longer reach large radii ($x=1$)
with a finite speed.

\begin{figure}
\plotone{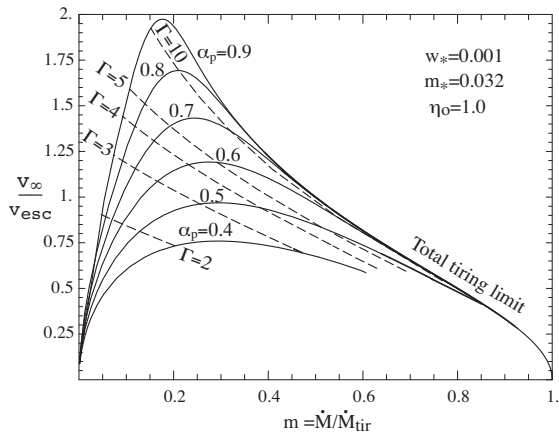}
\caption{
Terminal speed over escape speed, plotted vs.~photon tiring number 
$m\equiv {\dot M}/{\dot M}_{tir}$, for models with 
various specific power indices over 
the range $\alpha_{p} = 0.4-0.9$ (in increments of 0.1).
The dashed curves show contours of the associated Eddington 
parameter, for the specific values $\Gamma=$~2, 3, 4, 5, and 10.
The convergence of models along the upper right bound represents the 
total tiring limit.
}
\end{figure}

Fig.~5 plots the ratio of terminal speed to escape speed vs.~
photon tiring number  $m\equiv {\dot M}/{\dot M}_{tir}$, 
for models fixed with the standard parameter set
$\eta_{o}=1$, $m_{\ast} = 0.032$, and $w_{\ast}=0.001$, 
but with various specific power indices over 
the range $\alpha_{p} = 0.4-0.9$ (in increments of 0.1).
The dashed curves show contours of the associated Eddington 
parameter, for the specific values $\Gamma=$~2, 3, 4, 5, and 10.
Note that for lower $\alpha_{p}$ and moderately large $\Gamma$ the 
tiring becomes substantial. 
Indeed, the convergence of models along the upper right bound represents 
the {\em total} tiring limit, for which the combination of kinetic and 
potential energy of the wind has exhausted the entire base luminosity $\Lstar$.
Hence,  models along this limit  have little or no remaining 
luminosity to be observable as radiation.

This illustrates a remarkable property of these power-law porosity models.
For small power exponents and moderately large Eddington parameters, 
they can  readily exhaust nearly the entire available luminosity in 
driving the mass loss.

\section{Discussion}

\subsection{Comparison with Previous Porosity Analyses}

Let us now consider how the above scalings compare with results from 
previous analyses of porosity-moderated mass loss.
Specifically, as noted above (cf.\  eqn.~\ref{Mdotpor3}), \citet{Shaviv01MNRAS} 
has cast the mass loss scaling in terms of 
a ``wind function'' $\wf(\Gamma)$ 
times the deviation of the luminosity from its Eddington limit value,
\beq
{\dot M} = 
 \wf [\Gamma] ~
{ \Lstar- L_{Edd} \over a_{\ast} c}  
\,  =   \wf [\Gamma] ~ 
\left ( {\Gamma - 1 \over \Gamma} \right ) 
{\Lstar \over a_{\ast} c}
\, .
\label{Mdotwf}
\eeq
As noted in \S 5.2, for the single-scale porosity model, 
the wind function $\wf [\Gamma] \approx (\Gamma+1)/\Gamma \eta$,
which therefore varies 
from 
$\wf \approx 2/\eta$ for the marginally super-Eddington case $\Gamma-1 \ll 1$ 
to 
$\wf \approx 1/\eta$ for the strong super-Eddington limit $\Gamma \gg 1$.
For comparison, \citet{Shaviv01MNRAS} considered two limiting cases, of either 
blobs in the optically thin limit or thick but vertically elongated, and a third
case of a two phase Markovian mixture.
For all cases, he estimated that the wind function would be of order 
unity, which is indeed our estimate here from the single-scale porosity 
model.

However, for porosity models with power-indices $\alpha < 1 $, we find 
here that there can be a further increase above this basal mass 
loss of the single-scale model.
For example, for the canonical case $\alpha_{p} = 1/2$, comparison
of (\ref{Mdotwf}) and (\ref{Mdpowpore}) implies the wind 
function scaling to be
\beq
\wf [\Gamma] = { 4 \Gamma \over \eta_{o} }  ~~ ; ~~ \alpha_{p} = 1/2 
\, .
\eeq
A key new feature of the power-law porosity model is consequently that this 
wind function increases with the Eddington parameter, and so now 
can become substantially larger than just the order unity value 
derived for the single-scale model, and also inferred from the analyses by 
\citet{Shaviv01MNRAS}.
Since even the basal mass loss rate (\ref{mdbasal}) for the single-model can be a few 
percent of the tiring limit (\ref{mdtir}), the additional increase for 
moderately large Eddington parameter $\Gamma > 4-5$ can bring the 
overall mass loss close to this limit.

In addition, of course, the mass loss scaling depends on the scaling 
of the porosity-length $h_{o}$, and this might also depend on $\Gamma$.
In particular, we have assumed above that the porosity length scales 
with the pure-gas-pressure scale height $H$;
but in a medium in which  there is a residual smooth component that puts a 
floor $k_{min}$ in the effective opacity (see \S\S 5.3-6.2), 
the effective gravitational scale height at large depth is instead given 
by $H/(1-k_{min} \Gamma)$.
Scaling the porosity length scales with this larger stratification
height could then reduce the mass loss rate by a factor 
$(1-k_{min} \Gamma )$, 
which in the possible case that $k_{min} \Gamma \ltwig 1$, could 
represent a substantial reduction.
This general effect was also discussed by \citet{Shaviv01MNRAS}.

\subsection{Comparison to Inferred Mass Loss of Giant Eruptions}

Let us next consider the conditions needed for this 
power-law-porosity model to reach the observationally inferred mass loss 
rates of giant eruptions of Luminous Blue Variable (LBV) stars like
$\eta$~Carinae.
Analyses of the resulting Homunculus nebula 
\citep{Smith02} 
indicate that during the roughly 20-year outburst from ca.\ 1840-60, 
{$\eta$~Carinae lost  $2-10 \, \Msun$, 
representing an average mass loss rate of $0.1-0.5 \, \msbyr$.
Moreover, the expansion of the Homunculus nebula is inferred 
to be in the range $v_{exp} = 500-700$~km/s.
If we characterize this expansion velocity as being some order-unity 
factor of the escape speed from the surface of origin, 
$\sqrt{w_{\infty}} \equiv v_{exp}/v_{esc}$, 
then the total luminosity associated with
this wind mass loss is
\beqa
L_{wind} &=& {\dot M}  v_{esc}^{2} (1+w_{\infty}) /2
\nonumber
\\
&=& 30 \times 10^{6} \Lsun ~ {\dot M}_{1} ~ {\Mstar/\Rstar \over \Msun/\Rsun} 
(1+w_{\infty})
\, ,
\label{lwind}
\eeqa
where ${\dot M}_{1} \equiv {\dot M}/(1 \Msun/{\rm yr})$.
For a solar value of mass to radius, 
$v_{esc} \approx 620$~km/s~$\approx v_{exp}$;
thus, assuming $w_{\infty} \approx 1$  in this case, we see 
from eqn.~(\ref{lwind}) 
that the observationally inferred range of mass loss for $\eta$~Carinae 
requires a total wind luminosity in the range 
$L_{wind} \approx 6-30 \times 10^{6} \Lsun$ \citep{Smith02}. 

Moreover historical observations suggest a radiative luminosity of 
roughly $L_{obs} \approx 20 \times 10^{6} \Lsun$ \citep{DH97}, 
implying a total luminosity of up to $50 \times 10^{6} \Lsun$!

To facilitate comparisons with these observational values, let us now
convert the above dimensionless model results into physical units.
The above standard dimensionless parameters $w_{\ast} \equiv 
(v_{esc}/a_{\ast})^{2} = 0.001$ 
and 
$m_{\ast} = v_{esc}^{2}/2 a_{\ast} c = 0.032$ 
imply specific dimensional values 
for the surface escape speed
\begin{equation}
    v_{esc} = 2 c m_{\ast} \sqrt{w_{\ast}} \approx 620 {\rm km/s} 
    \approx v_{esc,\odot}
    \, ,
\end{equation}
and the surface sound speed
\begin{equation}
    a_{\ast} = 2 c m_{\ast} w_{\ast} \approx 20 {\rm km/s} 
    \, .
\end{equation}
The dimensional value of the terminal speed is thus given by 
$v_{\infty} =  620\,$km/s$~ \sqrt{w[1]}$.
From eqn.~(\ref{mdgens}), the dimensional mass loss is obtained from
\begin{equation}
\Mdot = 3.9 \times 10^{-3} \, {\Msun \over {\rm yr}} \, 
{\rho_{\ast} \over \rho_{o}} \, {M_{100} \over \eta_{o} a_{20}} 
\, ,
\end{equation}
where $M_{100} \equiv \Mstar/100 \Msun$ and 
$a_{20} \equiv a_{\ast}$/(20~km/s).

\begin{figure}
\plotone{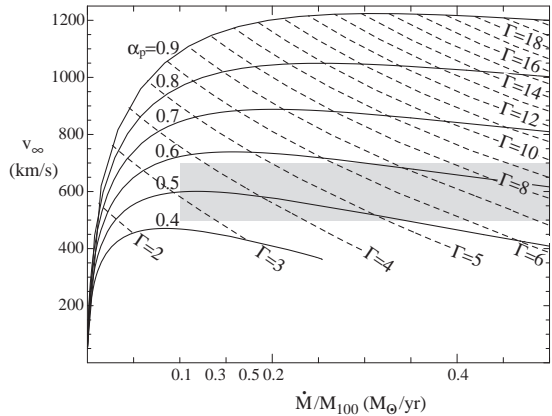}
\caption{
Same models as
fig.~5,
but now for corresponding dimensional values, 
showing terminal speed vs.~mass loss rate (scaled by $M_{100}$, 
the stellar mass in units of 100~$\Msun$).
The dashed curves again show contours of the Eddington 
parameter, now for integer values in the range $\Gamma=$~2 - 20.
Both plot sets assume the same dimensionless parameter combination,
$w_{\ast}=0.001$ and $m_{\ast} 0.032$, which together imply 
dimensional parameters of $a_{\ast} = 20$~km/s for the base sound speed 
and $v_{esc}=$~620~km/s ($= v_{esc,\odot}$) for the surface escape 
speed.
The semi-transparent box denotes the observationally inferred parameter range
for the 1840-60 giant eruption of $\eta$~Carinae that gave rise to the 
Homunculus nebula.
}
\end{figure}

Fig.~6 plots the terminal speed $v_{\infty}$ vs.~the mass loss rate 
$\Mdot$ for the same models shown in the dimensionless plots of fig.~5, 
for various power indices $\alpha_{p}$.
The dashed curves again show contours of the associated Eddington 
parameter $\Gamma$.
The observationally inferred ranges of mass loss and terminal speed
are denoted by the shaded box.
Assuming a canonical stellar mass $M_{100} =1$, 
note that even the upper range of mass loss, $\Mdot_{1} \approx 0.5$,
can be obtained with $\alpha_{p} \approx 0.6$ and $\Gamma \approx 8.5$.

However, note also from eqn.~(\ref{lwind}) that the associated wind
luminosity for this case is $L_{wind} \approx 30 \times 10^{6} \Lsun$.
From the conversion $\Gamma \approx 0.27 L_{6}/M_{100}$, this
implies an associated ``wind Eddington parameter'',
$\Gamma_{wind} \approx 8.1$, which therefore means that a fraction
$8.1/8.5$, or about 95\%, of the source luminosity is consumed in the 
wind!
This in turn implies that the observable radiative luminosity in such a model 
would be a mere 
\beqa
L_{rad,\infty} 
&\approx& 
(\Gamma-\Gamma_{wind}) L_{Edd} 
\nonumber
\\
&\approx& 
(8.5-8.1) \, 3.7 \times 10^{6} \Lsun
\nonumber
\\
&\approx &
1.5 \times 10^{6}\Lsun
\, ,
\eeqa
which is much less than the historically 
observed value of $L_{obs} \approx 20 \times 10^{6} \Lsun$
noted above \citep{DH97}.

Our further analysis indicates that fitting this observed luminosity 
together with the inferred mass loss and terminal speed requires a 
{\em reduction} of the assumed surface {\em escape speed}.
As shown in fig.~2, without photon tiring the ratio of terminal speed 
to escape speed increases with increasing Eddington parameter $\Gamma$.
This is a direct result of the fact that radiative driving scales 
with $\Gamma$, and so as seen from the equation of motion without tiring 
(eqn.~\ref{3.9}) the scaled acceleration likewise increases with $\Gamma$.
But the mass loss also increases with $\Gamma$, and in models 
with photon tiring, the increased tiring number reduces the driving
(see eqn.~\ref{eomportir}).
Thus, as shown in figs.~4 and 5, the terminal speed peaks and then 
decreases with higher $\Gamma$.
This makes it possible to obtain both a large mass rate and a modest 
terminal speed that is equal to the assumed escape speed. 
But because of the dominant role of tiring, a by-product of such 
models is that there is little remaining radiative energy to be
observed as emergent luminosity.

However, if we assume a lower escape speed, then models with less tiring 
can produce the observed terminal speed.
For example, taking the same stellar mass $M_{100}=1$ but 
an increased stellar radius $R_{100} = 3$ implies a reduced 
effective escape speed $v_{esc} = 360$~km/s, so that the same 
target expansion speed now requires 
a speed ratio $v_{\infty}/v_{esc} = \sqrt{3} \approx 1.73$, and 
therefore an energy ratio $w_{\infty}=3$.
For the targeted mass loss $\Mdot_{1}=0.5$, we then see from eqn.~
(\ref{lwind}) that the associated wind luminosity is now 
$L_{wind} \approx  20 \times 10^{6} \Lsun$.
Together with the observed luminosity, this requires of total base 
luminosity of $\Lstar \approx 40 \times 10^{6} \Lsun$, corresponding 
to an Eddington parameter $\Gamma \approx 10.8$.

By taking this $\Lstar$ with $M_{100}=1$ and $R_{100}=3$, and
keeping the remaining three parameters the same 
($\eta_{o}=1$, $\alpha=0.6$, $a_{20}=1$),  we then find that the
observable radiative luminosity is now higher than before
($L_{obs} \approx 13.4 \times 10^{6} \Lsun$), but still below the 
target value, essentially because of too much tiring from the 
still-too-high mass loss rate ($\Mdot_{1} \approx 0.76$).
To reduce the mass loss, we can increase the porosity parameter 
$\eta_{o}$ and/or the sound speed $a_{20}$.
Choosing the former, we find that  
$\eta_{o} = 1.56$ gives results within 2\% of the targeted values for all 
three observational parameters.

Of course, given that the model has a total of six adjustable parameters, the 
significance of being able to fit three observed values should not be exaggerated. 
Nonetheless, it is encouraging that this power-law porosity model can
reproduce such rather extreme mass loss conditions with quite 
plausible choices of these parameters.
It is particularly interesting that, because of the scalings of mass loss 
and flow speed with the Eddington parameter, and their interplay with 
photon tiring, reproducing both the inferred mass loss and radiative 
luminosity requires that the source star's surface escape speed is somewhat
smaller than the inferred expansion speed.

\subsection{Gravity Darkening and the Shaping of LBV Nebulae}

For simplicity, the above formalism has assumed a spherically 
symmetric mass loss, but in the case of $\eta$~Carinae, the resulting
Homunculus nebula exhibits a distinctly bipolar, prolate form.
Detailed analysis
\citep{Smith02}
indicates the nebula has both a higher speed and greater density toward 
the bipolar symmetry axis, suggesting that the 1840-60 giant outburst 
was itself intrinsically bipolar.
Moreover, there is evidence, from both 
ground-based interferometric observations \citep{vanBoekeletal03}
and from HST slit-spectra \citep{Smithetal03a}, 
the {\em present-day} wind of  $\eta$~Carinae is also bipolar about the same 
symmetry axis as the Homunculus.

One promising scenario, first proposed by \citet{OG97} 
and developed further by \citet{MM00}, \citet{MD02}, and \citet{DO02},
is that such bipolar mass loss is the natural consequence of 
radiative driving from a nearly critically rotating star with 
a substantial equatorial {\em gravity darkening} \citep{vZ24}.
However, since this basic idea arose from models of relatively low-density, 
{\em line-driven} winds \citep{OCG96}, the applicability to much denser, 
continuum-driven models for $\eta$~Carinae or other LBVs has been uncertain.
Eventually, the issue should be examined through 2-D dynamical models that 
account for the latitudinal mass and radiation transport within the 
optically thick, bipolar expansion.

But for now, it is worth noting that the underlying scalings 
for the speed and mass flux have some key similarities in both the 
basic line-driven-wind theory and the present models for
porosity-moderated continuum-driving of a dense, optically thick flow.
In particular, comparison of eqns.~(\ref{mdcak}) and 
(\ref{Mdpowpora})-(\ref{Mdpowpore}) shows that the mass loss rate in
both cases scales with the stellar luminosity times a correction 
factor that is a function of the Eddington parameter, i.e., 
$\Mdot \propto \Lstar f[\Gamma]$.
Accordingly, in a rotating star, the local surface mass flux 
${\dot m}_{\ast}(\theta)$ at any colatitude $\theta$ should vary in 
proportion to the local surface radiation flux $\Fstar (\theta)$, 
times a function of the local effective Eddington parameter 
$\Gamma \equiv \Fstar/g_{\eff}$, where $g_{\eff}$ is the effective, 
centrifugally reduced surface gravity.
But for the standard \citet{vZ24} gravity darkening 
scaling that $\Fstar (\theta) \propto g_{\eff} (\theta)$, we see that
the Eddington parameter is {\it latitudinally constant}, implying
then that the surface mass flux should scale simply as
\begin{equation}
    {\dot m}_{\ast}(\theta) \propto \Fstar(\theta) \propto g_{\eff}(\theta)
\, .
\label{msgeff}
\end{equation}
Since both the radiative flux and effective gravity are maximum at the 
rotational pole, eqn.~(\ref{msgeff}) shows that the mass flux should be
strongest near the poles.

Similar arguments can be made for the latitudinal variation of the 
flow expansion speed.
Again, the detailed results may depend on latitudinal components of 
the mass flow, radiation flux, and radiative force, and should be 
eventually be analyzed through 2-D models.
But within the context of simple 1-D scaling relations for both 
line-driven and porosity models, the outflow speed should follow 
the approximate scaling,
\beq
v_{\infty} (\theta) 
\propto
v_{esc} (\theta) 
\propto
\sqrt{g_{\eff}(\theta)}
\, .
\label{vinfot}
\eeq
Since the effective gravity is highest toward the poles, we can 
expect the nebula expansion to be faster near the symmetry axis.
Observations do indeed show that $v(\theta)$ for the polar lobes
of the Homunculus nebula is roughly proportional to the simple latitudinal
variation of escape speed from a rotating star \citep{Smith02}.

Overall, the expected faster polar flow speed can explain the
generally prolate form of the expanding nebula, 
while
the higher polar mass flux can explain the observationally inferred 
mass concentration near the polar symmetry axis.
Thus, an attractive feature of the porosity-moderated continuum-driven 
formalism is that it preserves these key 1-D flow scalings from line-driven 
models, while allowing extension to much larger mass loss rates.
However, as noted, more complete 2-D models should be developed 
to examine how these general scalings might be affected by latitudinal mass 
and radiation transport.

Finally,  while several LBV nebulae show such a bipolar 
or ellipsoidal form,  none except maybe HR~Carinae is as severely
pinched at the equator as $\eta$~Carinae. 
P Cygni -- the only other Galactic star observed to have a giant eruption 
approaching the extremity of that in $\eta$~Carinae -- 
has a nearly spherical ring nebula
\citep{NC97}.
In this context, it should be emphasized that the basic porosity model 
developed here  can accomodate a range of geometric forms, depending on the 
degree of rotation of the source star.

\section{Concluding Summary}

Let us conclude with an itemized summary of the goals, methods, and 
results of the analyses in this paper.
\begin{enumerate}
\item
The overall goal is to develop a radiative-driving
formalism for explaining the extremely large mass loss rates and moderately
high outflow speeds inferred in giant outbursts of Luminous Blue Variables, 
most particularly the 1840-60 outburst of $\eta$~Carinae that resulted in the
Homunculus nebula.
The general approach combines and builds upon two previous models for
radiative driving, namely the well-established CAK line-driving 
generally applied for more quiescent phases of hot-star wind mass 
loss, and recent notions of porosity-moderated continuum driving in 
stars that exceed a generalized Eddington limit 
\citep{Shaviv98, Shaviv01MNRAS}.
\item
As a basis for this synthesis, we first review the CAK line-driven 
formalism, showing thereby that for reasonable values of the 
line-opacity normalization, the expected wind momentum 
(i.e., the product of mass loss rate and flow speed) 
in such models falls well below what is inferred in LBV giant outbursts.
The formal divergence in CAK mass loss rate as a star approaches the 
Eddington limit is accompanied by a vanishing terminal speed, and 
moreover in practice is limited by the ``photon tiring'' of the 
finite luminosity available to lift material from the 
gravitational potential at the stellar surface.
\item
In the deep stellar interior, approaching and exceeding the Eddington 
limit leads to convective energy transport that lowers the radiative flux 
and allows a normal hydrostatic stratification with outwardly declining 
pressure and density.
In near-surface layers where the lower density makes
convective transport inefficient, a super-Eddington condition again 
implies that the outward radiative force exceeds gravity.
Hydrostatic stratification then requires an outward {\em inversion} of
pressure and density, which however can only be maintained over a 
limited range, given the inherent outer boundary condition of 
vanishing pressure and density.
\item
Moreover, any outflow initiated where convection first becomes 
inefficient would imply a huge mass loss rate, of order 
$\Lstar/a_{\ast}^{2}$, where $a_{\ast}$ is the sound speed;
this exceeds by a large factor
the ``tiring limit'' mass loss, $\Mdot_{tir} = 2 \Lstar/v_{esc}^{2}$,
for which the energy expended to lift material from the surface 
gravitational potential equals the assumed stellar luminosity.
The flow stagnation associated with photon tiring, together with other 
instabilities, seems likely to impart a complex, time-dependent, 3-D 
structure to the atmosphere of any star that approaches or exceeds the 
Eddington limit.
\item
As first noted by \citet{Shaviv98}, the ``porosity'' of such a 
structured, super-Eddington medium can lower the effective driving 
opacity in the deeper, denser layers, allowing a base hydrostatic balance 
that transitions to a supersonic outflow as the structures become optically 
thin, and hence are again subject to the full radiative force.
\item
We introduce a simple formalism that characterizes this effect in terms 
of a ``porosity length'' $h \equiv L^{3}/l^{2}$, where $l$ and $L$ 
represent the size and separation of individual clumps or blobs.
We show that a simple ansatz (somewhat analogous to the mixing length 
parameterization of convective transport) -- that this porosity 
length is an order-unity factor $\eta$ times the gravitational scale height 
$H$ -- leads to a mass loss rate 
$\Mdot_{por} \approx (1-1/\Gamma^{2}) L_{\star}/\eta a_{\ast} c$,
a scaling that 
is very similar to that derived previously by 
\citet{Shaviv01MNRAS}
for both vertical elongation and Markovian mixture models 
of atmospheric structure.
Though large, such mass loss is still typically only a few percent of 
the tiring limit.
\item
In analogy with the CAK formalism of driving by a power-law ensemble of 
lines, we then generalize to a ``power-law-porosity'' treatment,
characterized by a power index $\alpha_{p}$ and a 
porosity-parameter $\eta_{o}$ for the most optically thick blob.
For $\alpha_{p} \gtwig 1$, the derived mass loss rates are similar to the 
single-scale model, but for $\alpha_{p}<1$ they lead to enhancement 
by factors that scale with the Eddington parameter as 
$\Gamma^{-1+1/\alpha_{p}}$.
For large $\Gamma$, small $\alpha_{p}$, and/or small $\eta_{o}$,
this can now lead to overall mass loss rates that approach the photon 
tiring limit.
\item
Without tiring, the wind velocity follows a canonical 
``beta-velocity-law'' form, with index $\beta \approx 1$, and
a terminal speed proportional to the escape speed times a factor that 
scales roughly with $(\Gamma-1)^{(\alpha_{p}+1)/4}$.
With tiring, the velocity law can become nonmonotonic, with a terminal
speed that decreases with increasing mass loss rate.
Models with a terminal speed less than or equal to the 
surface escape speed are strongly tired, 
and so have a greatly reduced observable radiative luminosity.
\item
Given stellar parameters $\Lstar$, $\Mstar$, and $\Rstar$, plus a 
base sound speed $a_{\ast}$, 
the mass loss properties of the power-law-porosity model depend on the 
power index $\alpha_{p}$ and porosity-length parameter $\eta_{o}$.
For quite reasonable values, the model can reproduce the
observationally inferred mass loss and speed 
for the giant eruption of $\eta$~Carinae, 
but matching also the historically estimated radiative luminosity 
during this epoch requires a modest surface escape speed, with ratio
of stellar mass to radius perhaps a third of the solar value.
\item
Though the magnitude of mass loss greatly exceeds what's feasible with 
line-driving, the porosity model retains the key scalings with gravity
and radiative flux that would give a rapidly rotating, gravity-darkened 
source star an enhanced polar mass loss and flow speed, 
as is inferred from the bipolar form of the Homunculus nebula in 
$\eta$~Carinae.
\end{enumerate}

Overall, the extended porosity formalism developed here provides a promising 
basis for self-consistent dynamical modeling of even the most extreme
mass loss outbursts of Luminous Blue Variables, namely those that, like the giant 
eruption of $\eta$~Carinae, approach the photon tiring limit.
But of course much further work will be needed to substantiate and 
quantify the basic ``porosity-length'' phenomenology proposed here.
Much in the same way that modern-day multi-dimensional simulations of 
stellar convection have been used to test classical mixing length phenomenology,
future work should focus on developing 2-D or 3-D radiation 
hydrodynamical simulations of the nonlinear evolution of atmospheric 
structure in stars near and above the Eddington limit.
A particular challenge will be to develop fast techniques for treating
the multi-dimensional, nonlocal radiation transport in such a medium.
Of course, until such more fundamental calculations are carried out, 
it is difficult to appraise the overall applicability of our porosity-length 
approach.
But in any case the basic phenomenology helps provide a motivation 
and conceptual framework for carrying out and interpreting such 
more fundamental and challenging radiation hydrodynamical simulations.

\vspace{0.2in}

\noindent{\it Acknowledgements.}
SPO acknowledges support from NSF grant AST-0097983. 
NJS acknowledges the support from ISF grant 201/02.
We thank K. Davidson, R. Townsend and N. Smith for helpful discussions and 
comments.


\end{document}